\documentclass[prb,preprint]{revtex4-1} 
\usepackage[utf8x]{inputenc}
\usepackage{amsmath,amssymb,amsfonts} 
\usepackage{graphics}                 
\usepackage{color}                    
\usepackage{soul}
\usepackage{graphicx, subfigure}
\renewcommand{\thesubsection}{\thesection.\Alph{subsection}}
\newcommand{\ave}{\operatorname{ave}}
\newtheorem{teo}{Theorem}
\newcommand{\Apr}{\ensuremath{\mathrm{Apr}}}
\newcommand{\Cpr}{\ensuremath{\mathrm{Cpr}}}
\newcommand{\PAM}{\ensuremath{\mathrm{PAM}}}

\begin{document}

\title{Interplay between Airy and Coriolis precessions in a real Foucault pendulum}

\author{N.N. Salva}%
 \affiliation{ CONICET-CNEA, Centro Atomico Bariloche, Av Bustillo 9,500, Bariloche,}
\affiliation{Centro Regional Bariloche, Universidad Nac. del Comahue, Quintral 1250, Bariloche.  E-mail: natalia.salva@yahoo.com.ar}
\author{H.R. Salva}
\affiliation{ Instituto Balseiro, Universidad Nacional de Cuyo,  Av Bustillo 9,500, Bariloche. E-mail: h.r.salva48@gmail.com}


\begin{abstract}
We study the precession of a Foucault pendulum using a new approach. We characterize the support anisotropy by the difference between the maximum and minimum periods of the pendulum along the principal axes of the support. Then we compute the total precession rate, taking into account both the Airy precession of a spherical pendulum and the Coriolis precession due to the Earth's rotation. To study the resulting motion we developed a calculation loop, period after period, which describes the movement of the oscillatory trajectory of the bob.
To test our model, we mounted a test pendulum of $480.3$ cm length and measured its periods and precession. The rate of precession is sensitive to the dimensions of the pendulum, the anisotropy of the support, and the initial conditions.
We find that for certain amplitudes the precession can stop entirely, while the pendulum continues to oscillate. It is also possible to obtain continuous precession at lower oscillation amplitudes. 
We give an upper bound for this critical oscillation amplitude. We close with a discussion of the implications of our findings for the design of Foucault pendulums used in demonstrations and lab experiments.

\end{abstract}

\maketitle

\section{Introduction}
When Leon Foucault realized that the Earth's rotation could be demonstrated with a pendulums, he started to build them. In January 1851 he built a 2 m long pendulum, and one month later he built one 11 m long and hung it in the Observatory of Paris.\cite{Foucault} 
Shortly after, he hung a 67 m pendulum in the Pantheon of Paris, showing a clear demonstration of the Earth's rotation through the constant change in the plane of oscillation of the pendulum. For this last pendulum, the apex of the trajectory shifted by nearly 2 mm in each period. This change is known as the precession of the pendulum, or Foucault precession. 

Since then many pendulums had been mounted to demonstrate the Earth's rotation in halls of universities, museums, etc. Even though the pendulum is associated with Foucault, he wasn't the first to study them. In the year 1660 Vincenzo Viviani, a disciple of Galileo, noticed the precession of the plane of oscillation in his pendulum, even though he did not connect this behaviour with the Earth's rotation.\cite{Sommeria}  The interest for this demonstration has not diminished and Foucault pendulums are still built to show this interesting phenomenon. 

Our goal in this paper is to develop a realistic model of the behavior of a Foucault pendulum.  This is not as simple as textbook models of a "simple pendulum" suggest. After Foucault, many people tried to build pendulums to demonstrate the Earth's rotation. They concluded that the pendulum should be as long as possible and the bob's mass as heavy as possible.\cite{Somerville,Sommeria} During these trials, they noticed that when launched, the bob trajectory starts out being nearly linear. After some time, the bob starts to move in an elliptical trajectory. The most  accurate model for the Foucault pendulum is an ideal spherical pendulum.  
In 1851, the astronomer G.B. Airy studied the behavior of an ideal spherical pendulum when the mass described a trajectory whose projection onto the horizontal plane was slightly elliptical.\cite{Airy} He found that this elliptical trajectory precesses in the same direction as the elliptical motion, with constant angular velocity. He also determined the time to complete one whole revolution. This motion will be referred to as Airy precession, although some authors call it ``precession of area''.\cite{Griffith,Olsson} Therefore, besides Foucault precession due to the Earth's rotation, there is an intrinsic precession of the pendulum that should be considered.

H. Kamerlingh Onnes, A.B. Pippard and R. Verrault, and others, developed models to describe the trajectory of the pendulum.\cite{KO,Rene2018,Rene2007,Rene2019,Pippard} An analysis of experimental perturbations has been reported in Ref. \onlinecite{Braginsky,Pippard}. The most common approach solves the equations of motion, resulting from the law of conservation of energy and angular momentum.\cite{KO,Shulz} Coriolis precession is introduced through the action of a pseudo-force in a rotating frame of reference, which yields sinusoidal eigenfunctions.  For A. B. Pippard and V. B. Braginsky, Airy precession is one of many perturbations in their analysis, and the final equations are complicated.\cite{Pippard,Braginsky} 

The variables involved show themselves in different states of the experiment. Besides Coriolis precession and Airy precession, various factors can modify the motion of the pendulum's bob, such as air currents and temperature fluctuations. When attempting to account for all these different factors, we encounter many valid concepts, often resulting in complicated solutions, which obscure rather than clarify the panorama. Our model is minimal, choosing to simplify the analysis by ignoring these other factors and instead including only Airy and Coriolis precession along with the anisotropy of the support.

Our model is neither purely theoretical nor phenomenological. It is meant to be used after building a pendulum and measuring the main parameters.  We can then use these measured parameters in the model to predict how the pendulum will move under a wide range of experimental conditions.  In Section II we review  Airy  precession and Coriolis precession of an ideal spherical pendulum. In Section III we point out the deviations of a real pendulum from this idealization, considering a real support that introduces an anisotropy to the attracting potential of the pendulum. We suppose that the support is the main source of anisotropy and neglect all other sources (air movements, temperature changes, etc).  Because of the anisotropy, Airy precession begins to play a main role. This is the most important effect that can obscure Coriolis precession.  

In Section IV we develop a simple model for the Foucault pendulum, taking into account only two characteristics: Coriolis precession and Airy precession introduced by the support anisotropy. We also employ the minimum number of variables: the characteristic oscillation periods of the support, the length and the oscillation amplitude of the pendulum. An iterative numerical analysis reveals that the initial release angle of the pendulum is an important factor that can affect the pendulum's behavior. In Section V we show how to use the model to predict the pendulum's behavior, and finally in Section VI we share our conclusions. 

We believe that this model could be used to improve the design of Foucault pendulums. It can reveal the best conditions for the observation of Coriolis precession: amplitude, length, weight and shape of the bob, etc. It can also be used to validate new designs: the type of support, an electromagnetic brake, or a method to ensure continuous oscillation.

\section{The ideal spherical pendulum}
\begin{figure}[h!]
 \includegraphics[height=6cm]{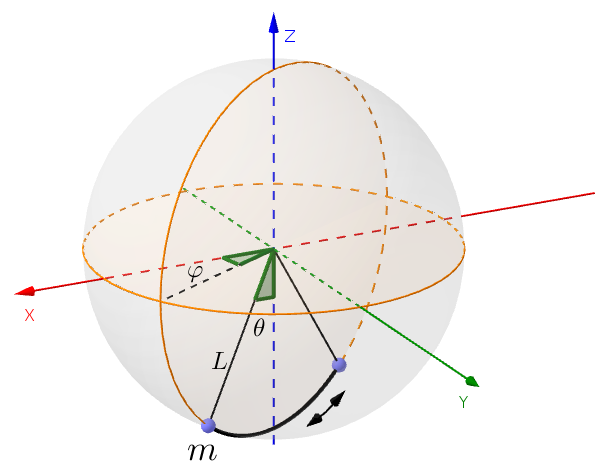}
 \caption{Spherical Pendulum of mass $m$ and length $L$.} \label{fig:pendEsferico}
\end{figure}

A Foucault pendulum is best represented by a spherical pendulum with small amplitude of oscillation ($\theta< 0.1$ rad). Let's assume we place the pendulum in a spherical coordinate system as shown in Figure \ref{fig:pendEsferico}, and  superpose a Cartesian coordinate system. 

The ideal spherical pendulum has the following properties:

\begin{enumerate}
\item Ideal support: There are no disturbances in the pendulum's motion produced by the support at the origin of the coordinate system. While the period of oscillation depends on $\theta$, it is independent of $\varphi$.
\item Point mass $m$: The bob is represented by a point mass $m$.
\item Non-extensible string: The connection between the support and the mass is a rigid segment. 
\item Constant amplitude: The pendulum does not lose energy during its motion, and therefore maintains a constant amplitude.
\item Uniform gravitational potential: The attractive gravitational potential is uniform. It has the same value at every point in space and is constant over time.
\end{enumerate}

We define the ``rest point'' of the pendulum as $(0,0,-L)$ in Cartesian coordinates, which represents the case of zero lateral force in a uniform gravitational field, where the potential energy is minimum.

The equations of motion follow from the law of conservation of energy and angular momentum. \cite{Griffith,Sommerfeld,MacMillan} The solution of these equations is complicated, and doesn't allow a direct analysis of the pendulum's behaviour. 

\subsection{Small Amplitudes}
If small amplitudes are considered and higher order terms are neglected, the solution is greatly simplified. The assumption of small amplitudes is equivalent to $\sin(\theta) \approx \theta$ and $\cos(\theta) \approx 1$. 
In this case
we get simple harmonic equations, and the projection of the path will be an ellipse centered on the rest point. This case will be discussed in the Section II.B. \cite{Griffith}

On the other hand, if we also assume that $\dot{\varphi}=0$, we are in the case of a simple pendulum (a degenerate ellipse). The mass will oscillate in a plane with constant $\varphi$ and the projection of its trajectory in the $xy$ plane will be rectilinear. In this case the period $T$ is equal to:

\begin{equation} \label{eq:idealperiod}
 T= 2\pi  \sqrt{\frac{L}{g}} \, ,
\end{equation} 

\noindent where $g$ is the gravitational acceleration, normally taken as $9.81$ m/s$^2$. We used the local value $9.80016$ m/s$^2$.\footnote{ This value was obtained as an average of several measurements done in the Physics Lab of the Balseiro Institute, Bariloche, Argentina.}

\subsection{Airy Precession} 
\label{ssec:airy}


\begin{figure}[ht!]
 \includegraphics[height=6cm,clip=true,trim=0 0 0 0]{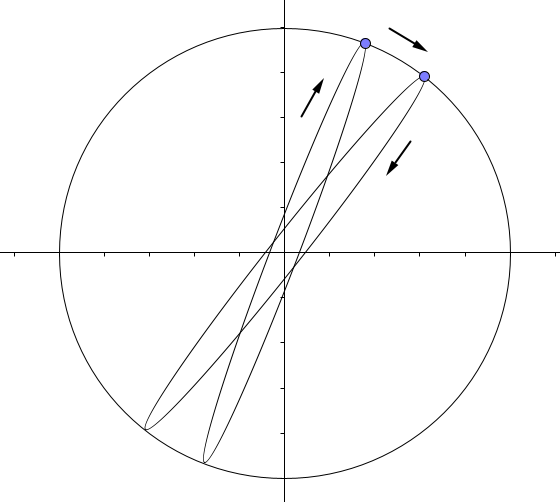}
 \caption{Bob's behaviour in the Spherical Pendulum with tangential impulse.} \label{fig:Airy}
\end{figure}


The astronomer G.B. Airy described the movement of an ideal spherical pendulum without any anisotropy, under the hypothesis that the bob describes an elliptical trajectory that differs very little from a straight line, as shown in Fig. \ref{fig:Airy}. \cite{Airy}  
He found that the major axis of the elliptical trajectory will slowly rotate around the rest point of the pendulum; that is, $\varphi$ will slowly change with time. This movement of the axis of oscillation is called ``precession''. 
The dimensions of the elliptical trajectory are constant during a complete turn, as there is no damping in this model. We can see in Fig. \ref{fig:Airy} the trajectory of the mass with an amplified precession for didactic purposes \footnote{Readers who wish to explore the motion of an ideal spherical pendulum may do so at https://demonstrations.wolfram.com/SphericalPendulum/}.  The precession
occurs because the period of motion around the ellipse is different than the time required
for the pendulum to swing back and forth.


If $L$ is the length of the pendulum and $a$ and $b$ are the major and the minor semiaxes of the ellipse, Airy showed that the time  ($T_a$) it takes the major elliptical axis to complete one revolution is:
\begin{equation}
T_a =\frac{8 L^2 T}{3 a b} \, [s] \qquad  \mbox{Period of Airy precession}
\end{equation}

\noindent where $T$ is the period of an ideal pendulum given by (\ref{eq:idealperiod}). The calculation of  $T_a$  was also performed in 1978 by M.G. Olsson in a simpler way using modern mathematics.\cite{Olsson} From this result we can calculate the Airy precession rate (\Apr{}),  the instantaneous angular velocity of $\varphi$, which can be calculated as $2\pi$ radians over the period $T_a$, using Eq. (\ref{eq:idealperiod}):
\begin{equation} \label{eq:Apr}
\Apr{} = \frac{2\pi}{T_a} =\frac{6 \pi}{8} \frac{ab}{L^2 T}=\frac{3 }{8} \frac{ab \sqrt{g}}{L^{5/2}}  \,  [rad/s] 
\end{equation}
Note that \Apr{} is proportional to $a$ and $b$, and inversely proportional to $L$. This illustrates why the pendulum should be as long as possible: to reduce Airy's perturbation.\cite{Somerville,Sommeria}

In addition, Airy concluded that the sense of rotation of the major axis of the ellipse is the same as the sense of rotation of the bob around the elliptical trajectory. \cite{Airy}
The precession rate depends on the dimensions of the elliptical trajectory, the length of the pendulum, and the period of oscillation, which, in turn, depends on the pendulum length and oscillation amplitude. 

Although the expression of equation (\ref{eq:Apr}) usually appears as a perturbation of the pendulum, to the authors' knowledge, the expression for $a$ and $b$ are never explicit.\cite{Olsson,Pippard}  In Appendix A, we derive expressions for $a$ and $b$ as functions of $\varphi$, $\omega$ and $t$ within our model.

\subsection{Coriolis Precession }
It is interesting to note how Foucault realized that a pendulum could be used to demonstrate the rotation of the Earth.
The path of the bob is generally a confusing matter. We have to establish two reference frames. One reference frame is fixed in space, or our planetary system (the fixed reference frame).
The other frame is fixed on the rotating Earth, whose angular speed is $\Omega_0 \approx$  15 $^\circ/$hr (the rotating reference frame).

In the fixed frame, if the pendulum is launched from its maximum amplitude directly towards the rest point, it will oscillate in a plane with constant $\varphi$ and the projection of its trajectory in the $xy$ plane will be rectilinear. This is the case of a simple pendulum, where the planar angular momentum (\PAM{}) in the $xy$ plane is zero. 
Now, if we consider the Earth's rotating frame and launch the pendulum in the same way, it will oscillate in a rotating plane and $\varphi$ will slowly change with time. 

Let's assume that we have an ideal spherical pendulum at the north pole, held with a support that coincides with the Earth's axis of rotation. The rest point is also located on the Earth's axis of rotation. 
The pendulum is released as it would be in a museum or lab on earth, without any azimuthal component of the launch velocity in the rotating frame. 
If we release the mass $m$ at an angle $\theta_0$ in a position $\varphi_0$, and angular velocity $\omega_0=0$ in the rotating frame, then in the fixed reference frame $\theta_F=\theta_0$, $\varphi_F=\varphi_0$, and its angular velocity is $\omega_F=\Omega_0$. The pendulum's path in the fixed reference frame will be an ellipse that never swings through the rest point, because there is an azimuthal component in its angular velocity. If we suppose that the Airy precession of the ellipse is much smaller than 15 $^\circ/$hr, then we can ignore this precession in the fixed reference frame.

But what is the trajectory as seen from the rotating reference frame of the Earth?

\begin{figure}[h!]
 \includegraphics[height=6cm]{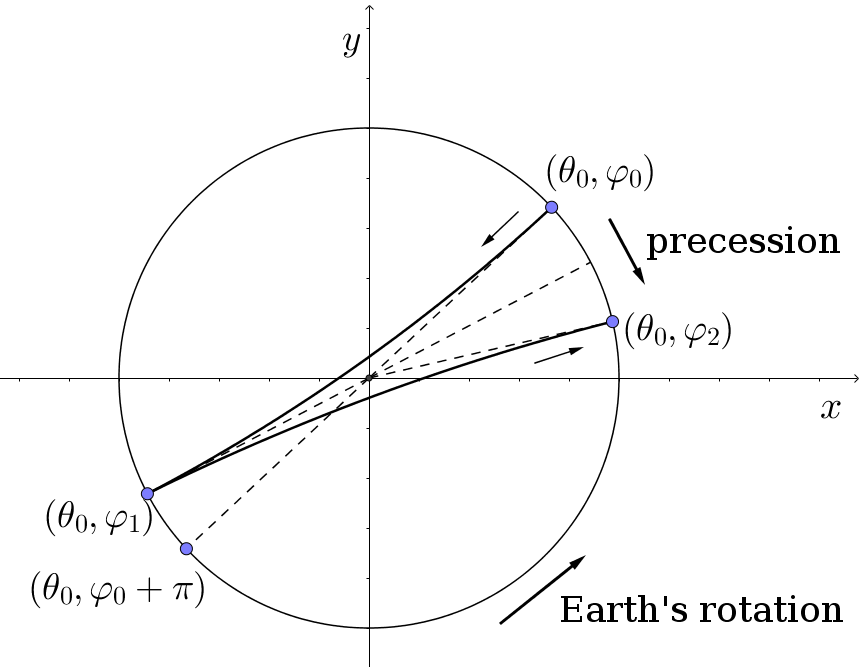}
 \caption{Bob's motion for a spherical pendulum at the North Pole, as viewed by an observer rotating with the Earth in CCW.} \label{fig:Estrella}
\end{figure}

Since the Earth rotates with an angular velocity $\Omega_0 \approx 15$ $^\circ/$hr CCW,  $m$ will also rotate with that angular velocity in the fixed reference frame so that $\omega_F = \Omega_0$ at the launch point. 
The \PAM{} in that state will be of $\PAM{}= k \theta_0 \omega_F = k \theta_0 \Omega_0$, which is constant due to conservation of angular momentum. The constant $k$ depends on the length of the pendulum and the mass $m$, among other parameters. Since \PAM{} is constant, it follows that when $\theta$ decreases, $\omega$ in the rotating frame will have to increase ($\omega>0$); consequently, $\varphi$ will grow. When approaching the rest point, $\omega$ will be maximum, and $\theta $ will have a minimum value since it cannot be zero. At that point, $\varphi \approx \varphi_0 + \pi/2$, and $m$ will not pass over the rest point. Then $\theta$ will increase until the mass reaches ($\theta_0$,$\varphi_1$), but $\varphi_1$ will be less than $\varphi_0 + \pi$, the azimuth corresponding to a rectilinear oscillation. The oscillation will continue and reach ($\theta_0$,$\varphi_2$) and $\varphi_2$ will be less than $\varphi_0$ . Figure \ref{fig:Estrella} shows the projection onto the $xy$ plane, describing the movement of the mass, in the rotating frame, which we referr to as the ``shining star path''.\cite{Somerville} After one oscillation, $m$ will have changed its launch coordinates from $(\theta_0, \varphi_0)$ to $(\theta_0, \varphi_2)$, where $\varphi_2 <\varphi_0$. The initial position will have rotated CW. (In the south pole the precession is CCW.) The oscillation will continue and consequently it will show the rotation of the Earth.

Foucault couldn't provide convincing arguments to explain the behaviour of his pendulum when it wasn't located at the north pole. It was Joseph Liouville (1809-1882), Professor of Mathematics at the College of France, who analyzed the problem from a vector point of view and determined  the Coriolis precession rate (\Cpr{})  for any latitude $\lambda$:\cite{Liouville}

\begin{equation} \label{eq:CorLat}
 \Cpr{}= - \Omega_0 \sin(\lambda),
\end{equation}

\noindent where $\Omega_0 \approx 15$ $^\circ/$hr is the angular velocity of the Earth. We adopt the convention that the latitude ($\lambda$) is negative in the southern hemisphere, and that it is positive in the northern hemisphere. Therefore, \Cpr{}  has the opposite sign as $\lambda$: that is, precession is CCW in the southern hemisphere, and CW in the the northern hemisphere.  Note that at the equator the latitude is equal to zero, an there will be no Coriolis precession. 

Subsequently, advances in mathematics allowed for the description of the Coriolis precession as a result of a pseudoforce, expressed through a cross product. The equation of motion corresponding to a rotating reference frame with a constant angular velocity $\boldsymbol{\omega}$ is:\cite{Goldstein, Somerville}
\begin{equation}
   m \boldsymbol{\ddot r}=\boldsymbol{T}+m \boldsymbol{g}-2m \, \boldsymbol{\omega} \times \boldsymbol{\dot r}, 
\end{equation}
where $m$ is the mass, $\boldsymbol{r}$ is the position vector, $\boldsymbol{T}$ the tension in the string, and $\boldsymbol{g}$ the vector gravity.
In this model, equation (\ref{eq:CorLat}) may also be derived, which illustrates the dependence of the Coriolis force on the latitude.

\section{The real pendulum}

In the real world, a bob is generally hung with a wire connected to a fixed support. There are many designs of supports. The easiest one is perhaps to weld a wire (such as a piano or guitar string) to a massive support.\cite{Longden} In this section we introduce the influence of the support to the equations of motion, and combine the effects of Coriolis and Airy precession.

The main differences between a real pendulum and the ideal spherical pendulum are the following:
\begin{enumerate}
\item A real support: The support produces disturbances in the pendulum's motion. The period depends on $\theta$ and also on $\varphi$.
\item Extended Mass: The mass is a sphere or an oblate ellipsoid with its major axis horizontal at the rest point (to minimize air resistance).
\item Extensible string: A wire may stretch when a mass is suspended.  However, its length does not change much during oscillation, and the assumption of a non-extensible string remains a good approximation in most cases.
\item Variable amplitude: There is loss of energy due to air resistance, energy transfer in the support-mass linkage, and friction in the support itself. This produces damping of the amplitude.
\item Nonuniform gravitational field: A uniform gravitational field is generally a good approximation. However, the attractive gravitational potential can change if the location of the pendulum is close to a large variable mass such as maritime tides.\cite{Rene2007}   
\end{enumerate}

There are three common methods of support:
a wire welded to a fixed mass or clamped with a chuck, 
a gimbal system with metal crosshead, 
and a rigid pendulum rolling on a ball, as described in references \onlinecite{Allais,Salva2013}. The third method has the drawback that the pendulum's oscillation plane cannot rotate 360 $^\circ$.
In any of these real pendulums, the mass is not a point and the construction of the pendulum may break rotational symmetry. (A wall-clock pendulum is an extreme example.)  Any anisotropy will be reflected in the period of oscillation of the pendulum. 
Therefore in a real Foucault pendulum we have to solve two important problems: the appearance of anisotropy in the period of the pendulum and a loss of energy, primarily to air resistance.

The problem of energy loss can be solved by a device that restores the energy lost in each cycle. Foucault, in 1851, had already designed (but not implemented) a mechanical device to keep his pendulum of the Pantheon in Paris oscillating.  Without such a device, losses drastically reduce its amplitude after 5 hours.\cite{Foucault2,Tobin,Tobin2} A modern solution is to place a magnet at the bottom of the bob and an accelerator coil at the rest point in the base of the pendulum, and control the system through an electronic circuit.\cite{Salva} With this device, the amplitude of the oscillation can be very stable.

The anisotropy in the period of the pendulum was first studied by Heike Kammerlingh Onnes in his Ph.D. thesis at the National University of Groningen (Netherlands) in 1879. \cite{KO} He constructed a rigid pendulum supported by a gimbal system with a crosshead, having two rotational axis at $90$ $^\circ$. He found that the period depended on the axis of rotation of the crosshead. He called this difference ``linear anisotropy". 
In general, for any kind of support, there will be a maximum period ($T_{\max}$) in one plane of oscillation and a minimum period ($T_{\min}$) in another plane of oscillation, for the same oscillation amplitude ($\theta_0$). This difference in periods accounts for all the sources of the anysotropy of the system. Therefore we can characterize any support anisotropy by the difference in the periods.  We define a dimensionless support time anisotropy as follows:
\begin{equation} \label{eq:SuppAn}
\Delta T= \frac{T_{\max} -T_{\min} }{ T_{\max} +T_{\min} }.
\end{equation}

We can also define the effective difference in length of the pendulum for the maximum and minimum periods using Eq. (\ref{eq:idealperiod}):
\begin{equation} \label{eq:deltaL}
\mbox{$\Delta L=\dfrac{g(T_{\max}^2-T_{\min}^2)}{4\pi^2}$ } 
\end{equation}
This $\Delta L$ describes the anisotropy in the support as an equivalent variation in the length of the pendulum. In the case of a welded wire, $\Delta L$ reflects the uniformity of the bending modulus with respect to $\varphi$. In the case of a gimbal, it indicates discrepancies between rotation axes, among other anisotropies. For a rigid pendulum rolling on a ball, it reflects the spherical precision of the small ball.

The first consequence of this period difference, is that there are two principal axes. These principal axes are defined as the axes where the period of oscillation reaches its maximum and minimum. These axes may be perpendicular (as for a gimbal with a metallic crosshead) or may have any angle between them (as for a welded wire). 

Another consequence of this anisotropy is that a linear trajectory of the mass will become elliptical, and Airy precession becomes important.
If we release the pendulum along a principal axis, Coriolis precession causes the major axis of the trajectory to precess away from the principal axis of the support. The linear trajectory then becomes an ellipse.  Its semiminor axis slowly increases, then decreases as the elliptical orbit rotates toward the other principal axis. The bob in its elliptical trajectory moves either counterclockwise (CCW) or clockwise (CW), depending on the phase difference between the main axes. 
When the orbit is aligned with the other principal axis, the minor axis of the ellipse tends to zero, and the trajectory becomes a straight line again. Afterwards it starts to describe an elliptical trajectory again, but the sense of rotation changes as it passes through the second principal axis. This change in the sense of rotation occurs every time the bob passes through one of the principal axes. (See Appendix B.)

\section{Motion in a real pendulum}

Now, we will look at the combined effect of Coriolis and Airy precession, which may reinforce or counteract each other.
Up to now, we have seen that if the bob moves in an elliptical trajectory, the plane of oscillation will precess with an Airy precession rate given in Eq. (\ref{eq:Apr}). We have also established that any support has two principal axes with two different periods of oscillation. 

The main conclusions of the previous section are the following.
\begin{enumerate}
\item The support has two extreme periods of oscillation (maximum and minimum) at different (but not necessarily perpendicular) azimuthal orientations.
\item The dependence of the period on the azimuth causes an initially linear trajectory of the pendulum to become an ellipse, which precesses with the Airy precession rate.
\item The bob's trajectory is a shining star path due to Coriolis precession modified by the elliptical path of Airy precession. 
\end{enumerate}

\subsection{Motion model}
Our convention for the precession angle $\varphi$  is that it increases when it moves CCW in the plane of the principal axes, and decreases when it moves CW. We assume here that the principal axes of the support are orthogonal. 
We introduce an orthogonal set of axes $xy$, aligned with the principal axes of the support, and we assume that $T_{\min}$ occurs along the $x$-axis and $T_{\max}$ occurs along the $y$-axis.

\begin{figure}[h!]
 \includegraphics[height=6cm]{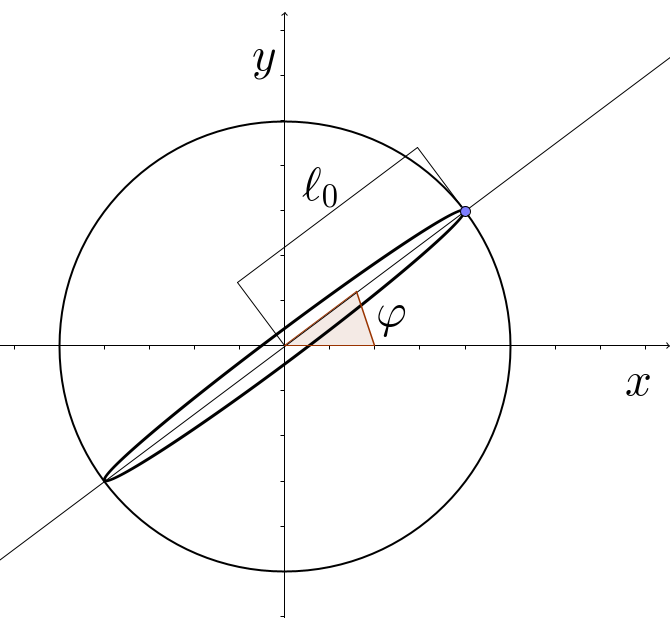}
 \caption{Motion model} \label{fig:ElipseModelo}
\end{figure}

We begin our analysis with the pendulum oscillating with its major axis at angle $\varphi$. Then we have the following equations, which describe the bob's trajectory of the pendulum in the rotating frame.

\begin{equation} \label{eq:motion0}
\begin{cases}
x(\varphi ,t) = \ell_0 \cos(\varphi ) \cos ( \omega_x t)\\
y(\varphi ,t) = \ell_0 \sin(\varphi ) \, { \cos} (\omega_y t)
\end{cases}
\end{equation}

Here, $\ell_0=L.\sin(\theta)$ is the oscillation amplitude, $t$ is the time variable, $\omega_x = 360/T_{\min} \,[^\circ/\mbox{s}]$ and  $\omega_y = 360/T_{\max} \,[^\circ/\mbox{s}]$ are the angular frequencies. We suppose that the support anisotropy is small ($T_{\max} \approx T_{\min}$); therefore $\omega_x$ will not differ too much from $\omega_y$. We define $\Delta \omega=\omega_x-\omega_y$ and $\omega=\omega_y$. Then we have:

\begin{equation} \label{eq:motion}
\begin{cases}
x(\varphi ,t) = \ell_0 \cos(\varphi ) \cos ( { \omega t+\Delta \omega t})\\
y(\varphi ,t) = \ell_0 \sin(\varphi ) { \cos (\omega t)}  
\end{cases}
\end{equation}

These equations describe an elliptical trajectory of the bob, in which the phase difference increases in time. As a result, the dimensions and orientation of the ellipse are constantly changing. Hence, the trajectory is not a Lissajous figure, where the components have a constant phase relationship.  Ref.~\onlinecite{Shulz} derives similar expressions but assumes a constant phase difference.

Another characteristic of this model is that the bob never passes over the rest point of the pendulum (the origin of the orthogonal reference system), unless it is oscillating along one of the principal axes.

As Coriolis (\Cpr{}) and Airy (\Apr{}) precessions are independent variables with completely different origins, we introduce them to our model as follows:

\begin{equation} \label{eq:varphipunto}
\frac{\partial \varphi}{\partial t}= \Cpr{} +\Apr{}   
\end{equation}

where $\Cpr{}>0$ for the southern hemisphere (CCW rotation), and $\Cpr{}<0$ for the northern hemisphere (CW rotation).

The sign of \Apr{} is positive if the bob moves CCW around its elliptical trajectory and negative if it moves CW.
In this model, the total precession of the pendulum is the sum of the individual precessions, with their corresponding signs determined by the quadrant ($\varphi$) and latitude ($\lambda$).  The Coriolis precession rate \Cpr{} only depends on the latitude. 
However, Airy precession reverses when it crosses a principal axis of the support, as Airy states in item 19 of Ref. \onlinecite{Airy}.
That is, in one quadrant the total precession rate is the sum of both precession rates ($\Cpr{}>0$, $\Apr{}>0$), and in the next quadrant it is the difference of the two precession rates ($\Cpr{} >0$ and $\Apr{} <0$). Therefore, $|\Apr{}|$ must be smaller than \Cpr{}, for the pendulum to continue precessing. Appendix B provides a detailed description of the sense of rotation of the ellipse.  

As the ellipse precesses  slowly, we can assume that
$\Delta \omega t$ is approximately constant during one period. Using this assumption, we can determine the axes of the ellipse. In Appendix A we give a detailed description of the solution using the Principal Axes Theorem for quadratic forms.\cite{Kolman} The resulting expressions are:

\begin{equation}
 \mbox{Major Semiaxis: }   a= \frac{\ell_0 }{\sqrt{2} } \, \sqrt{1+\sqrt{1-\sin^2(2 \varphi) \sin^2(\Delta \omega t)}}\label{eq:Maxe}     
\end{equation}

\begin{equation}
 \mbox{Minor Semiaxis: }
  b= \frac{ \, \ell_0   \,|\sin(2 \varphi) \sin(\Delta \omega t)|}{\sqrt{2}\sqrt{1+\sqrt{1-\sin^2(2 \varphi) \sin^2(\Delta \omega t)}}} \label{eq:Min}
\end{equation}

The term $\Delta \omega$ is a direct consequence of the anisotropy in the pendulum support. The bigger the anisotropy, the bigger is $\Delta \omega$.\\

To study the {precession of the major axis} we developed an iterative numerical algorithm. We define the average period as $T_{\ave}=(T_{\max} +T_{\min})/2$. For each iteration of the numerical algorithm, we determine the ellipse dimensions $a$ and $b$, \Apr{}, $\Delta \varphi=( \Cpr{}+ \Apr{})T_{\ave}$,  and update $\varphi$.  We define as $T_x$ and $T_y$ the periods at the $x$ and $y$ axes. Since $T_{\min}$ was along the $x$-axis and $T_{\max}$ along the $y$-axis, we have $T_x<T_y$. \\

The loop begins with the initial pendulum parameters:  $L$, $\ell_0 $, $T_x$ ,$T_y$ and $\varphi_1$. We define $n$ as the iteration number and initialize at $n=1$.  The stop criteria of this loop is when $\Delta \varphi$ is smaller than $10^{-4} \, [^\circ]$,  which means the pendulum has stopped precessing. The stop criteria can be changed for other criteria, such as a certain number oscillations.\\

\fbox{ \parbox{0.97\linewidth}{ 
Do:
 \begin{enumerate}
    \item Calculate: $t_n = n \cdot T_{\ave}$, and $ \Delta \omega \cdot t_n$
    \item Calculate: major semiaxis $a$, minor semiaxis $b$, and \Apr{} using Eq. (\ref{eq:Maxe}), (\ref{eq:Min}) and (\ref{eq:Apr}), respectively.
    \item Calculate: $\varphi_{n+1} = \varphi_{n} + (\Apr{}+\Cpr{}) \cdot T_{\ave}$.
    \item $n=n+1$.
 \end{enumerate}   
 While  $(\varphi_{n+1} - \varphi_{n})>0.0001 \, ^\circ$.
}}\\

In some pendulums, it is possible for precession to stop, if it oscillates for long enough.  If the dimensions and sense of rotation of the elliptical trajectory ever become such that $\Apr{}=-\Cpr{}$, this will produce a total precession equal to $\Cpr{}+\Apr{}=0$. In this case, the pendulum will oscillate in an elliptical trajectory without precessing for a long time. We define $\varphi_s$ as the azimuth when the precession stops.

\section{Results}

In this section we apply the calculation loop for different pendulums. We begin with an experimental pendulum of $480.3$ cm length ($479$ cm is the center mass length) and $30$ cm maximum amplitude, briefly described in Appendix \ref{secc:pend}. It was constructed in the southern hemisphere at a latitude of $-40.8$$^\circ$, yielding $\Cpr{} =9.8^\circ/\text{hr}$.  The mass was hung from a gimbal with a metallic crosshead, featuring a stabilized amplitude system, and had periods of  $T_x=4.3946$ s and $T_y=4.3948$ s. We measured these periods using the ``microsec()'' function of an Arduino UNO R3 board. These values were averaged over 14 periods (approximately one minute), which was possible because the pendulum precesses only 0.16 $^\circ$ per minute. 
We then used these oscillation periods to model the pendulum's behavior for different amplitudes. (In reality, these periods depend on the amplitude and differ by about $0.1$ ms.) 
Table \ref{tab:1} shows the calculated results for our experimental pendulum.  Table \ref{tab:2} shows the results for a hypothetical pendulum with a length of L=1000 cm and a difference in oscillation periods of 0.1 ms.

\begin{table}[h!]
\begin{center}
 \begin{tabular}{|c|c|c|c|c|}
         \hline \hspace*{0.4cm} Amplitude \hspace*{0.4cm} & \quad \Apr{}$_{\max}$  for $\varphi<90 ^\circ$  \quad &  \hspace*{1cm} $b_{\max}$ \hspace*{1cm} & \quad $100 \cdot a_{\min} / a_{\max}$ \quad & \hspace*{1cm} $\varphi_s$ \hspace*{1cm} \\ 
          (cm)&  ($^\circ/$hr)&   (cm) & \% &  ($^\circ$) \\ \hline
         30  & 71.6 &  5.05    & 98.6 & 92 \\ \hline
         20   &  42.2 &   4.5 & 97.4 & 93.4 \\ \hline
         10  &  15.6  &   3.5  & 93.8 & 103.4 \\ \hline
         \end{tabular}
\end{center}

    \caption{Pendulum with $L= 479$ cm, $T_x=4.3946$ s, $T_y=4.3948$ s,  $\Delta T= 2.2 \cdot 10^{-5}$, and $\Delta L$ = 0.43 mm.}
    \label{tab:1}
\end{table}

\begin{table}[!h]
\begin{center}
    \begin{tabular}{|c|c|c|c|c|}
         \hline \hspace*{0.4cm} Amplitude \hspace*{0.4cm} & \quad \Apr{}$_{\max}$  for $\varphi<90 ^\circ$  \quad &  \hspace*{1cm} $b_{\max}$ \hspace*{1cm} & \quad $100 \cdot a_{\min} / a_{\max}$ \quad & \hspace*{1cm} $\varphi_s$ \hspace*{1cm} \\ 
          (cm)&  ($^\circ/$hr)&   (cm) & \% &  ($^\circ$) \\ \hline
         50&  16.31&   4.27&  99.63 &  97.50\\ \hline
         30&  7.45&  3.26&  99.41 &105.12\\ \hline
         20 &   3.76&  2.47 &99.23 & 118.74\\ \hline
         18 &  3.12&  2.28& 99.19  &125.34\\ \hline
    \end{tabular}
    \end{center}
   \caption{Pendulum with $L= 1000 \mbox{ cm}, T_x=6.3436 \, \mathrm{ s}, T_y=6.3437 \, \mathrm{ s}$, $\Delta T= 7.9 \cdot 10^{-6}$, and $\Delta L$ = 0.31 mm.}
    \label{tab:2}
\end{table}

Tables \ref{tab:1} and \ref{tab:2} show, for different oscillation amplitudes, the maximum Airy precession rate for the first quadrant, the maximum value of the minor semiaxis, and the minimum value of the major semiaxis (as a percentage of its maximum), during the time it takes the precession to stop. We also show the azimuth $\varphi_s$ of the pendulum when it stops precessing. In these calculations, the pendulum was released from $\varphi_1=1 \,^\circ$ in a quadrant where \Apr{} adds to \Cpr{}. In the next quadrant ( $\varphi > 90$ $^\circ$), $\Apr{}<0$ and the precession stops when \Cpr{} equals $-\Apr{}$. Consequently, the stopping azimuth is always greater than 90$^\circ$.
The major semiaxis experiences a small decrease (always less than 1 per cent) and the extreme values of both the major and minor semiaxes occur in nearly the same azimuth. 

Using Eq. (\ref{eq:Apr}) we can replace $a$ and $b$ by Eq. (\ref{eq:Maxe}-\ref{eq:Min}), bounding $|\sin(2 \varphi)|$  and   $|\sin(\Delta \omega t)|$ by 1, and get an upper bound for \Apr{}:

\begin{equation}
 \Apr{} \leq \frac{4.86 \times 10^5}{\sqrt{2}}\frac{\ell_0\,^2}{ L^2 T} 
 \qquad [^\circ/\mbox{hr}]\label{eq:Apr_ub}
\end{equation}

This gives the Airy precession rate in degrees per hour.  The units which must be used are $\ell_0$ [cm], $L$ [cm], and $T$ [s] (which can be replaced by $T_{\ave}$ [s]).  Note that the maximum value for both sine functions may not occur simultaneously, but they occur approximately at the same angle.  From this expression, we can calculate an upper bound on $\ell_0 $ for which $\Apr{} < \Cpr{}$:

\begin{equation} 
 \ell_0 \, ^2 < \frac{\sqrt{2} }{4.86 \times 10^5} \Cpr{} L^2 T\label{eq:amp_osc}
\end{equation}
The value of \Cpr{} in [$^\circ/$hr] should be used here.  This bound can be used to guarantee a continuous precession after building a pendulum.

\begin{figure}[ht!]
    \centering
    \includegraphics[height=6.3cm,clip=true, trim=70 10 70 50]{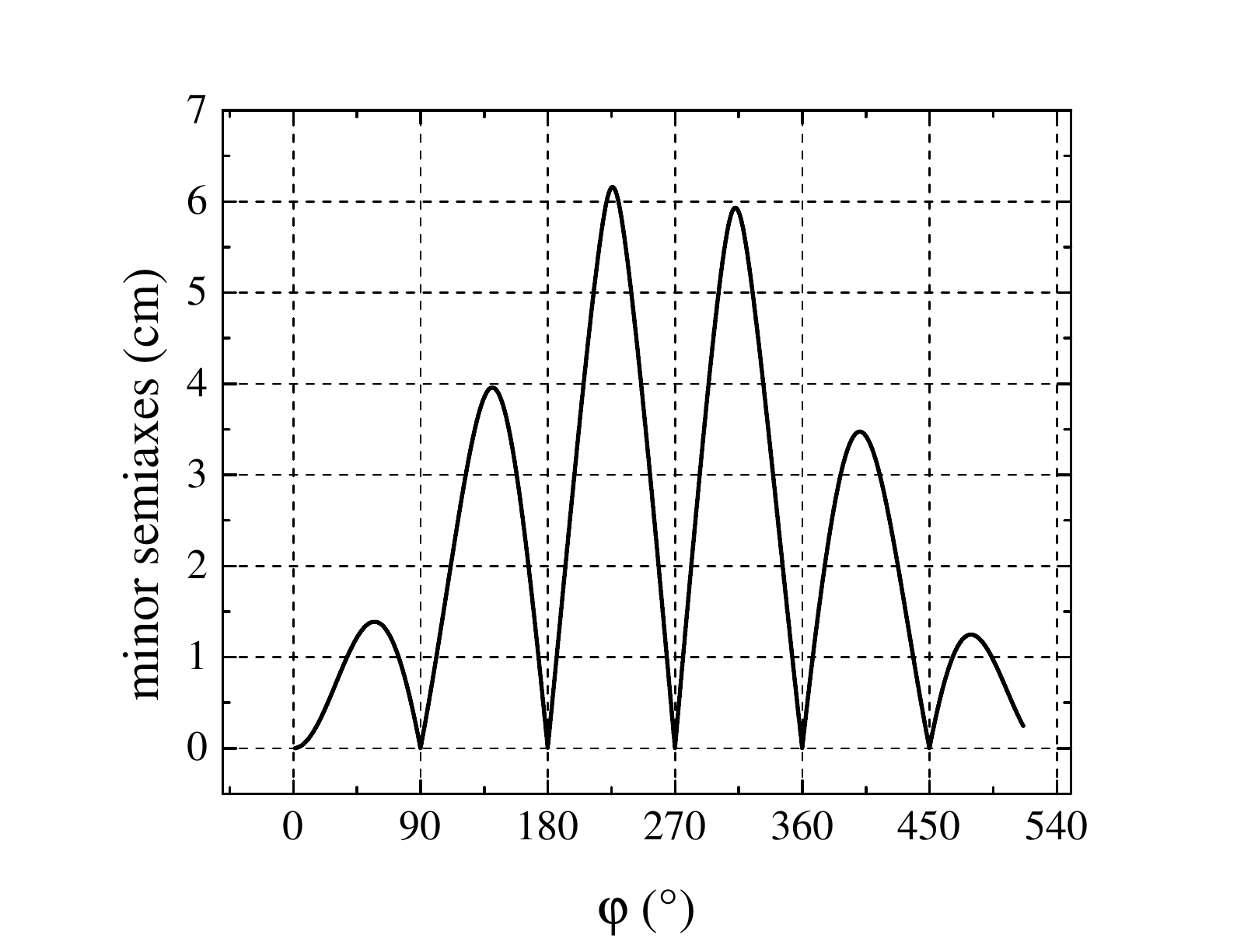} \includegraphics[height=6.3cm,clip=true, trim=40 10 70 50]{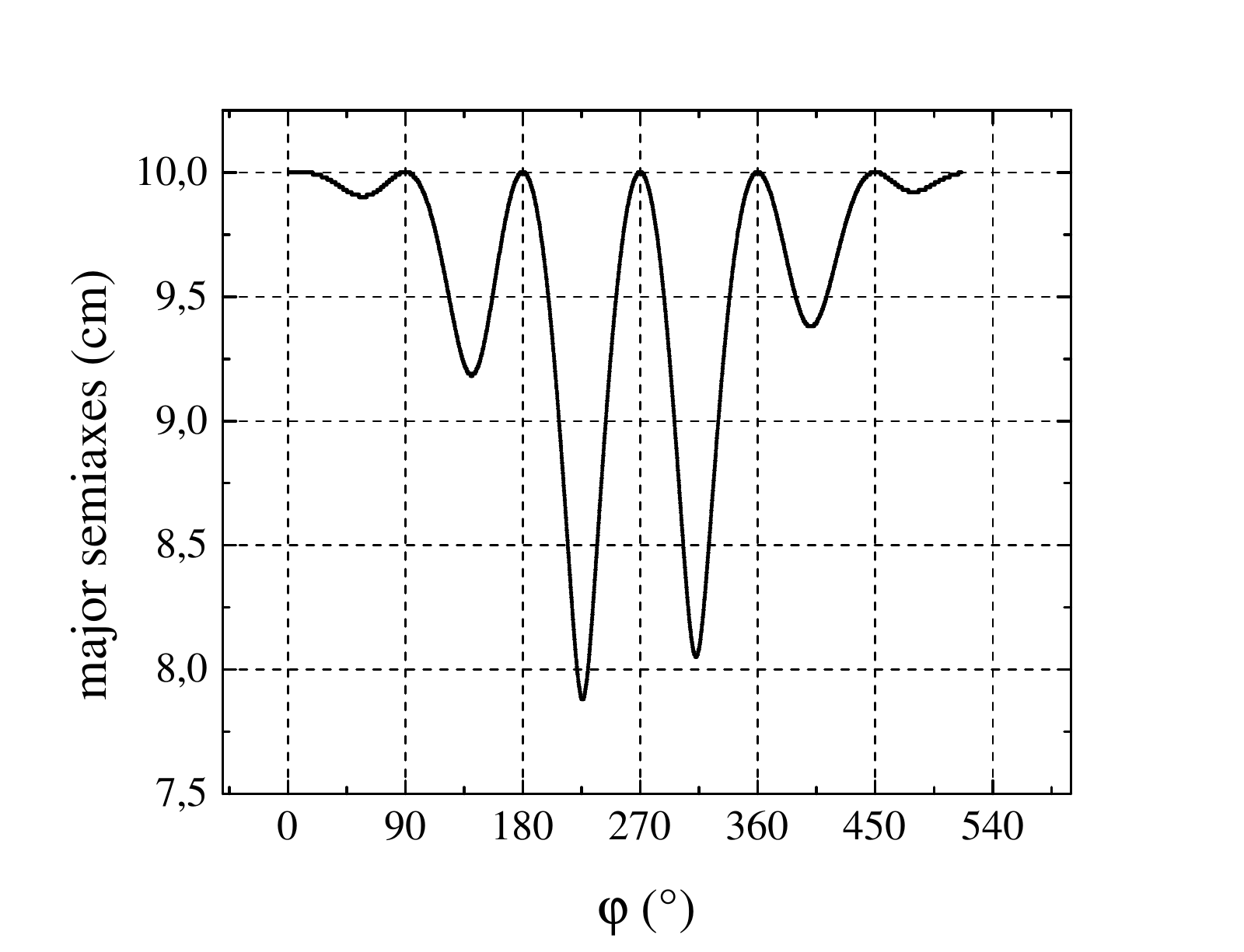} \\  
    (a) \hspace*{7cm} (b) \\
    \includegraphics[height=6.3cm,clip=true, trim=60 10 70 50]{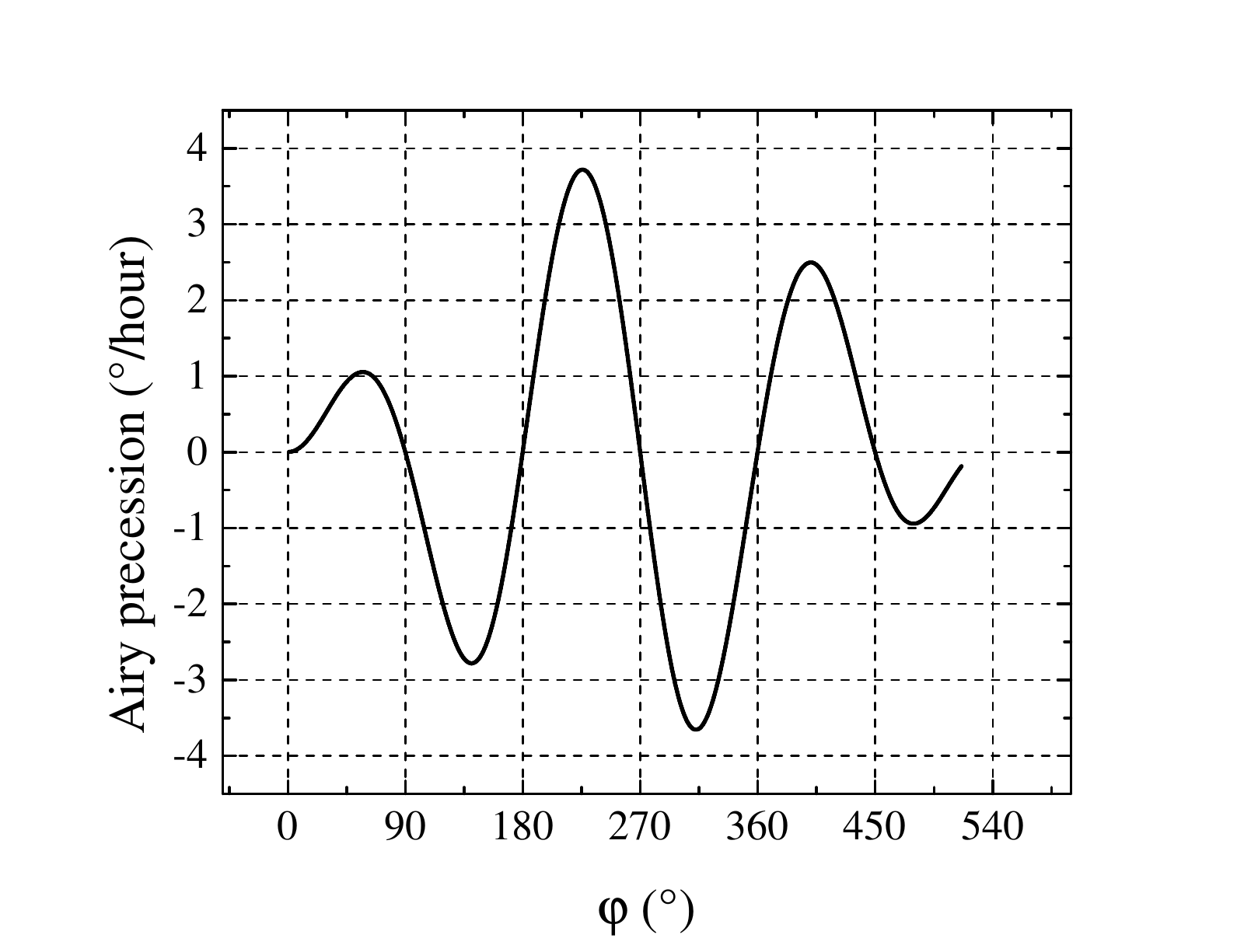}  \includegraphics[height=6.3cm,clip=true, trim=50 10 70 50]{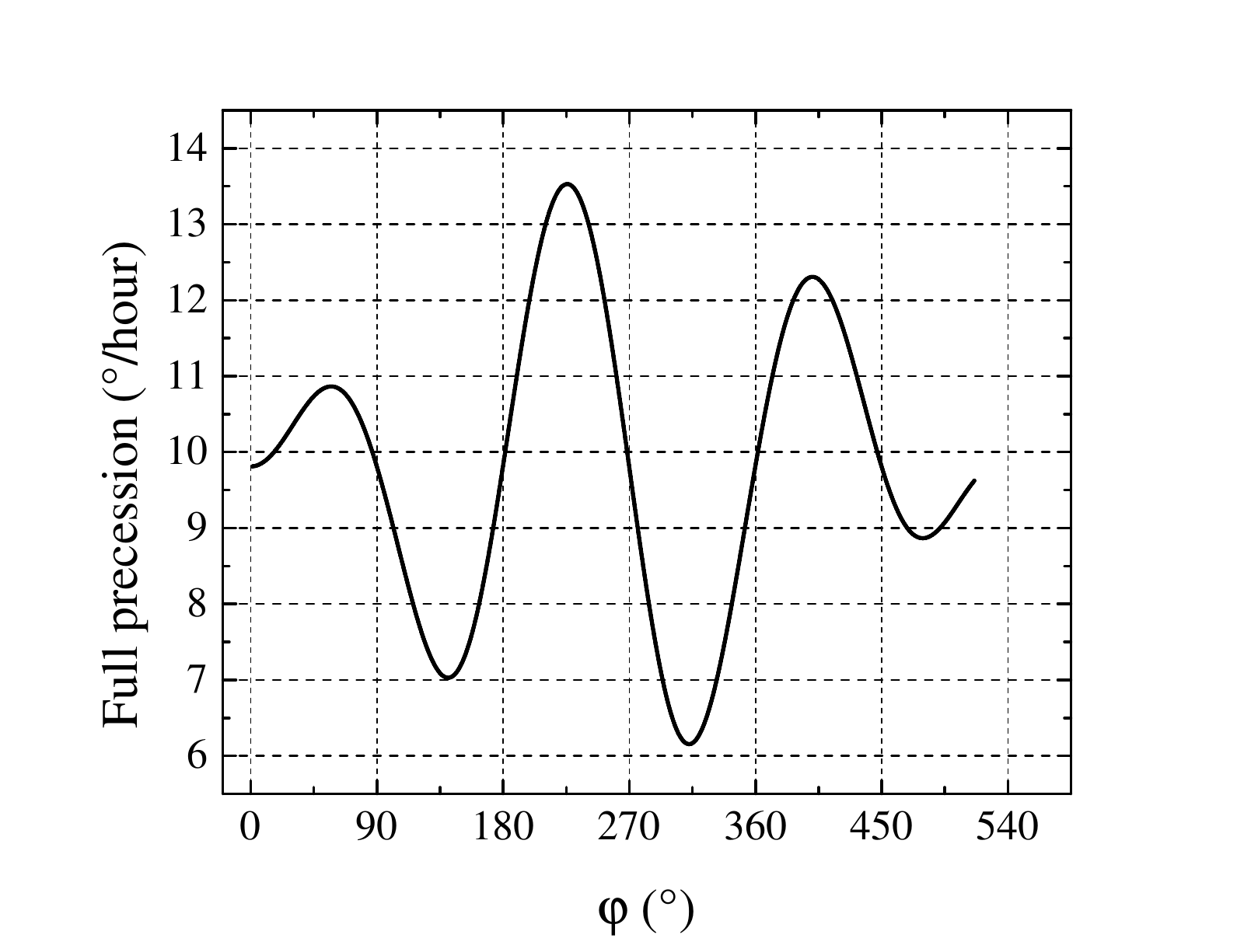} \\  
    (c) \hspace*{7cm} (d) \\
    \caption{Pendulum with $L= 1000 \mbox{ cm}, T_x=6.3436\mbox{ s}, T_y=6.3437$ s, $\Delta T = 7.9 \cdot 10^{-6}$, and oscillation amplitude of 10 cm. (a) Minor Semiaxis. (b) Major Semiaxis. (c) Airy precession rate. (d) Full precession rate.}
    \label{fig:2}
\end{figure}

In the following examples, precession does not stop, and the stop criterion for the calculation loop is a maximum number of iterations. (We considered up to 40000 iterations.) 
In Fig. \ref{fig:2} we show the minor and major semiaxes, the Airy precession rate and the combined precession rate, for a pendulum of length 10 m and oscillation amplitude 10 cm, assuming a time anisotropy of the support of $0.1$ ms.
We can see the effects of $\sin(2\varphi)$ and $\sin(\Delta \omega t)$ in Eqs. (\ref{eq:Maxe}) and (\ref{eq:Min}) on the minor and major semiaxes.
In particular, the minor axis is equal to zero every 90 $^\circ$. This directly affects \Apr{}, which also goes to zero every 90 $^\circ$. The term $\sin(\Delta \omega t)$ affects the  maximum of the minor axis and the minimum of the major axis. The maximum of the minor axis increases for $0<\Delta \omega t<90$ $^\circ$, and decreases for $90^{\circ} < \Delta \omega t < 180^{\circ}$. The opposite happens for the major semiaxis.
Also note that we calculated the motion for more than one whole revolution ($\varphi>360^\circ$). During the first rotation, the pendulum's precession does not stop, because the Airy precession rate is too small.

To study how the initial azimuth affects the behavior of the pendulum, we applied the algorithm to a pendulum with a length of 67 m and an oscillation amplitude of 3 m, similar to the one built by Leon Foucault in 1851 hung in the Pantheon of Paris. We consider it hung in the southern hemisphere with latitude of $-40.8$ $^\circ$ as before. The support time anisotropy is assumed to be equal to $3 \cdot 10^{-7}$ which leads to $\Delta L=$  0.08 mm .

\begin{figure}[ht!]
    \centering
  \includegraphics[height=6.5cm,clip=true, trim=70 10 70 50]{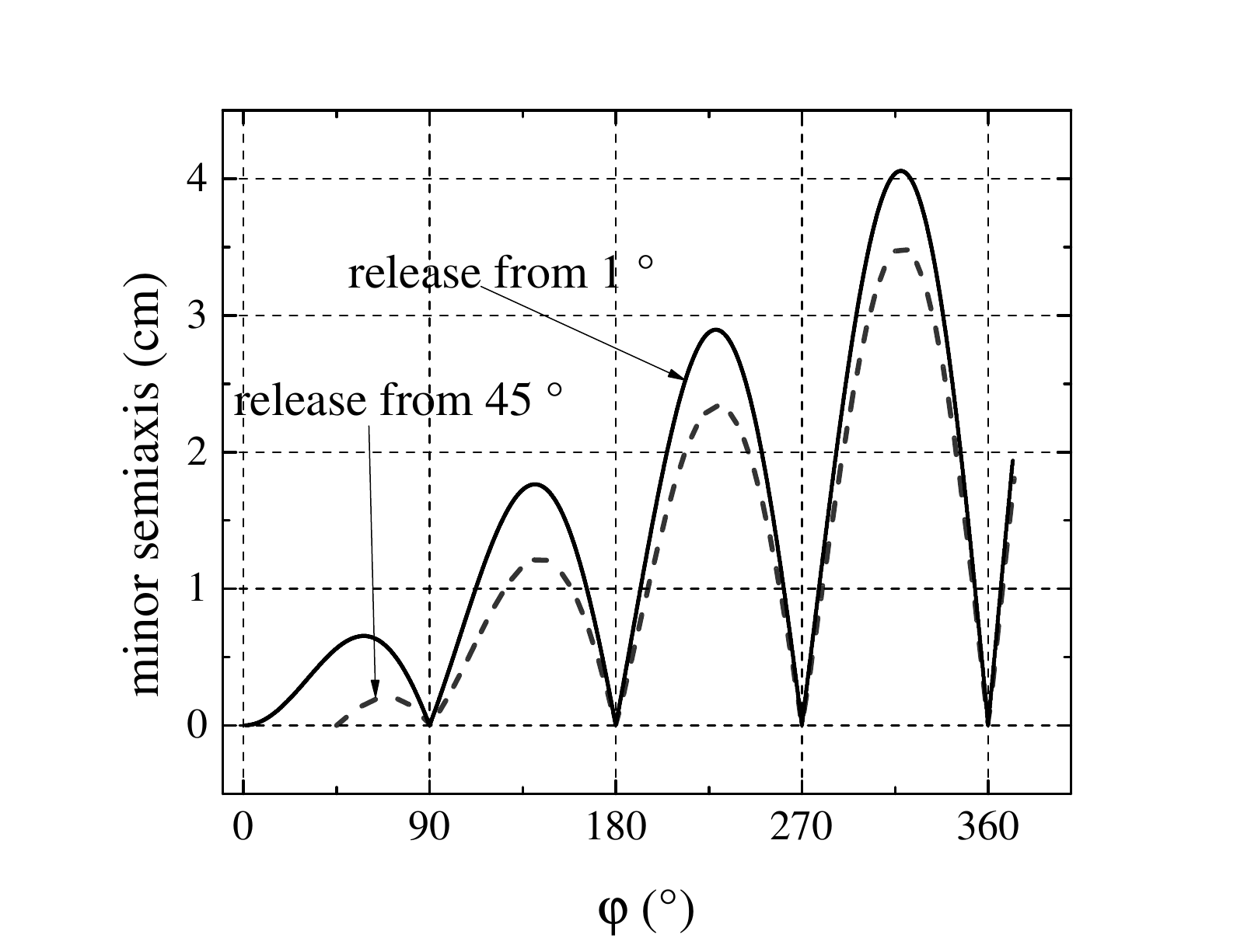}   \includegraphics[height=6.5cm, clip=true, trim=40 10 70 50]{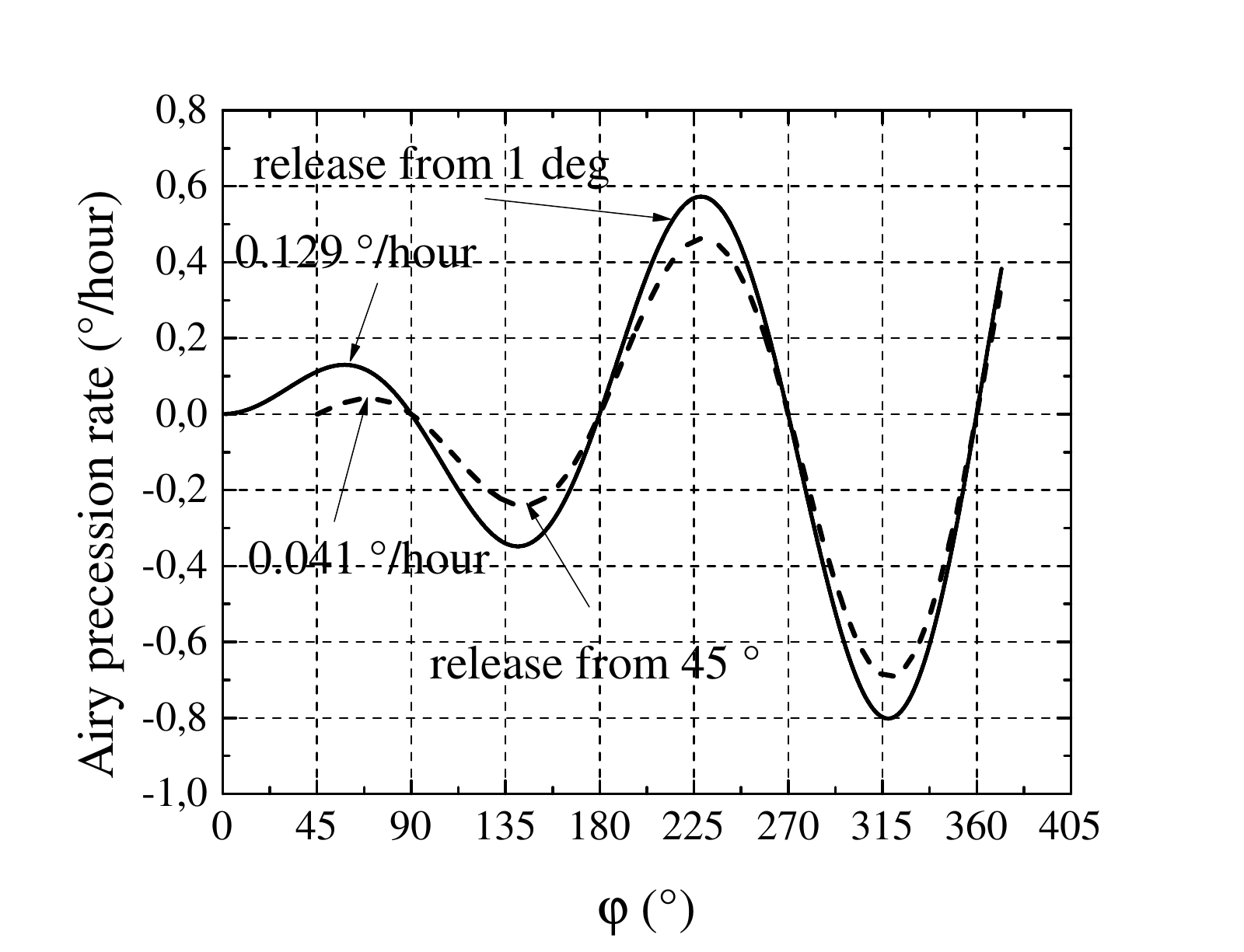}\\
    (a) \hspace*{7cm} (b) \\
    
    \caption{Pendulum with $L=67$ m, $\ell_0 =3$ m, $T_x=16.41999$ s, $T_y=16.420$ s, and $\Delta T= 3 \cdot 10^{-7}$, similar to Foucault's pendulum in the Pantheon of Paris. (a) Minor Semiaxis. (b) Airy precession rate.}
    \label{fig:1}
\end{figure}

In Fig. \ref{fig:1} we show the minor semiaxis and the Airy precession rate, computed using our algorithm. The effect of $\sin( \Delta \omega t)$ is responsible for the continuous growth of the minor semiaxis.
In Fig.\ref{fig:1}(b), the phase $\Delta \omega t$ at {the azimuth of }360 $^\circ$ was approximately 4 $^\circ$. If we let the pendulum continue to oscillate, the phase will  grow, as in Fig. \ref{fig:2}, until \Apr{} reaches the same value as \Cpr{} ($9.8 ^\circ/$hr), and therefore the pendulum {precession} will stop about at 840 $^\circ$.

In Fig. \ref{fig:1}(a), if the initial angle is $\varphi_1=45^\circ$, the minor axis grows more slowly than if we released it at $\varphi_1= 1^\circ$, near the principal axis. Therefore, the Airy perturbation is smaller, as seen in Fig. \ref{fig:1}(b). 

\begin{figure}[ht!]
    \centering
    \includegraphics[height=6cm,clip=true, trim=10 10 70 50]{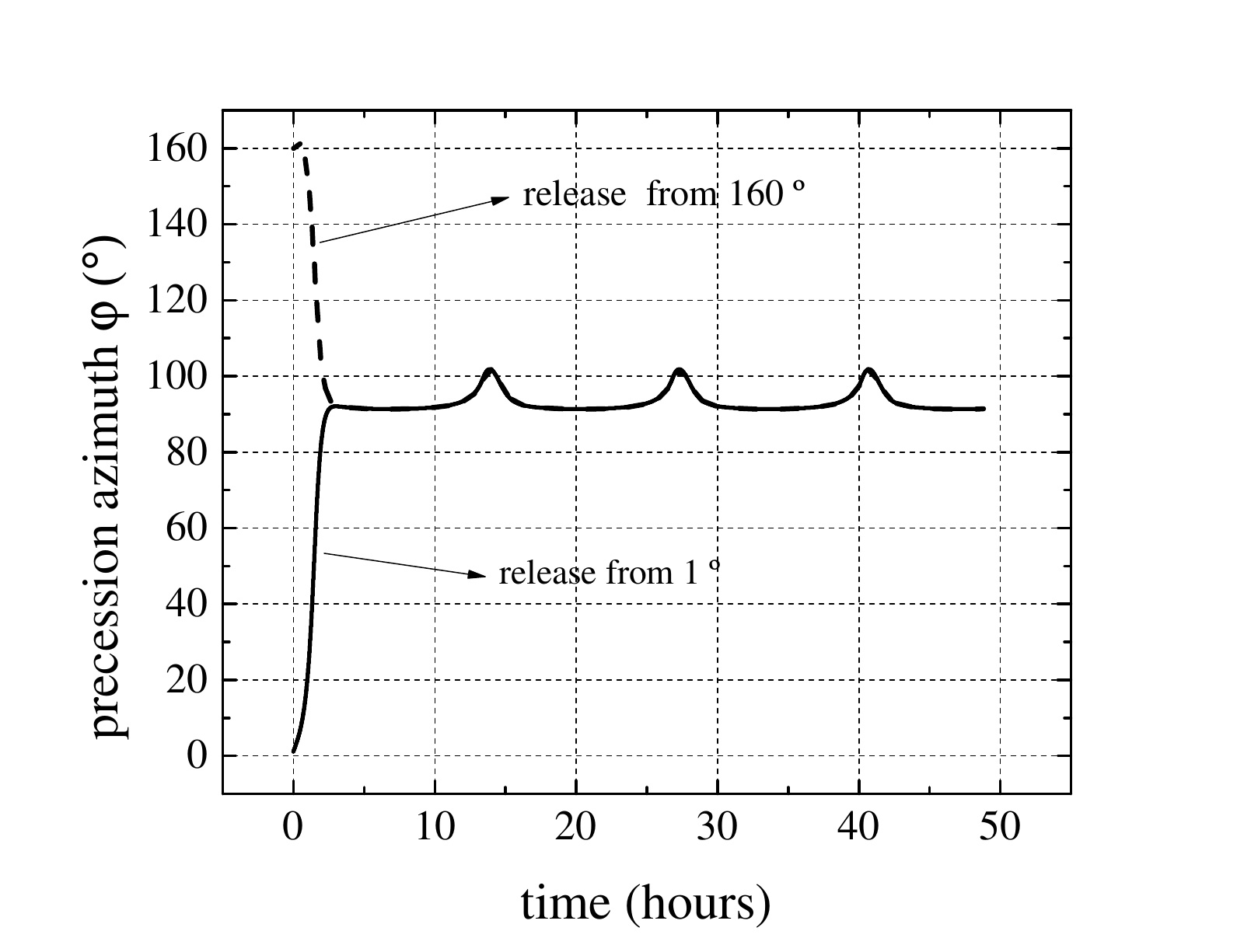}  \includegraphics[height=6cm,clip=true, trim=10 10 70 50]{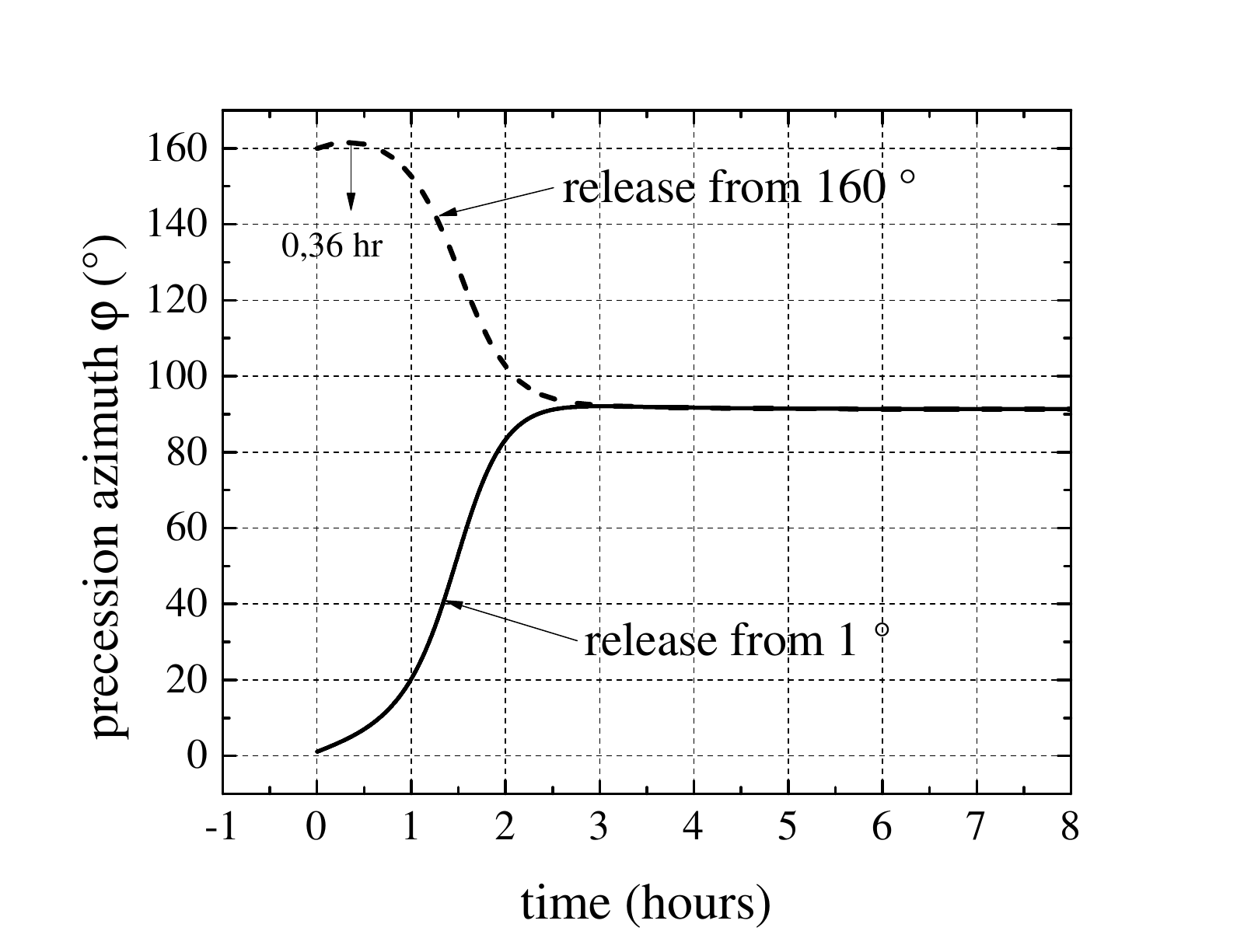} \\  
    (a) \hspace*{7cm} (b) \\
    \includegraphics[height=6cm,clip=true, trim=10 10 70 50]{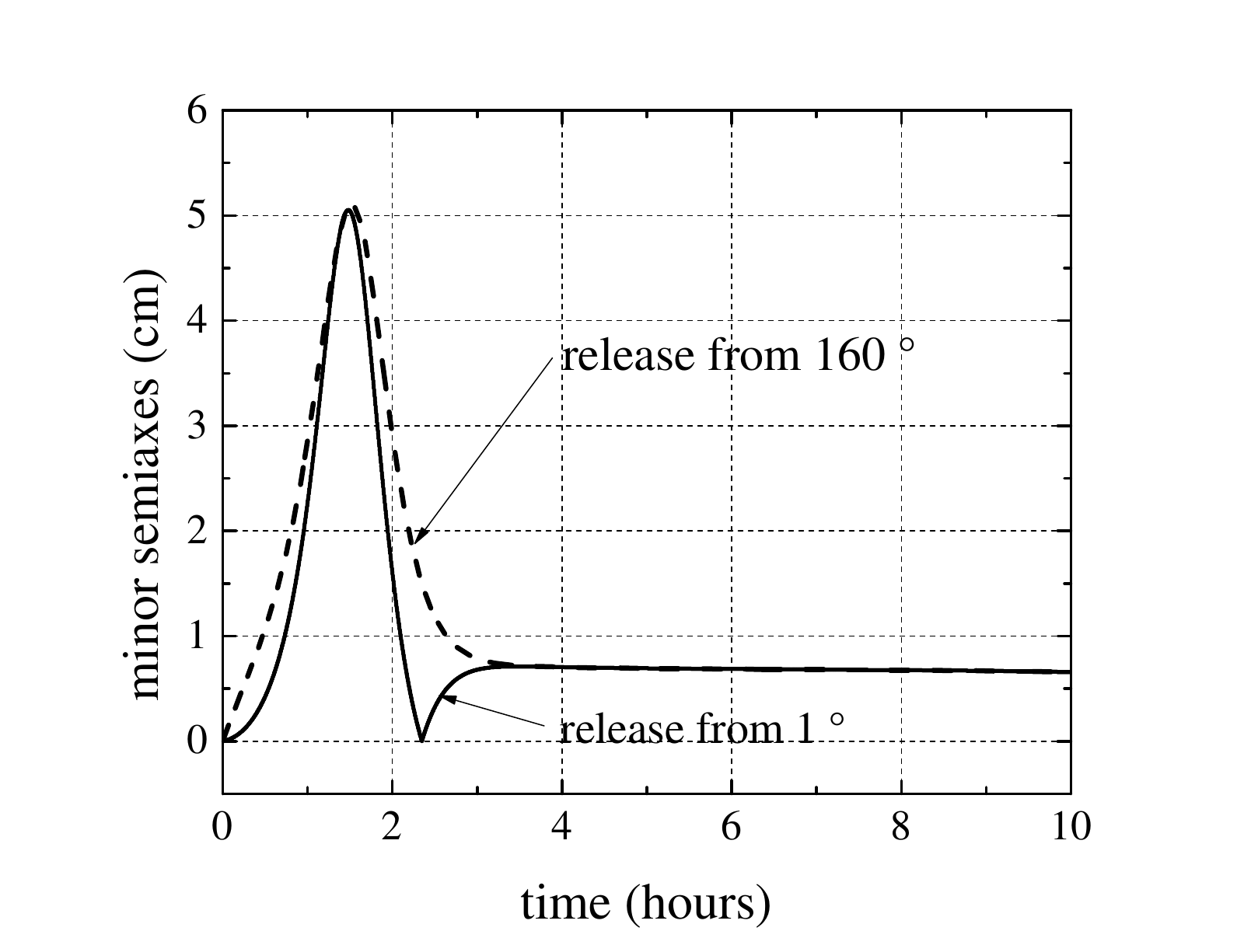}  \includegraphics[height=6cm,clip=true, trim=10 10 70 50]{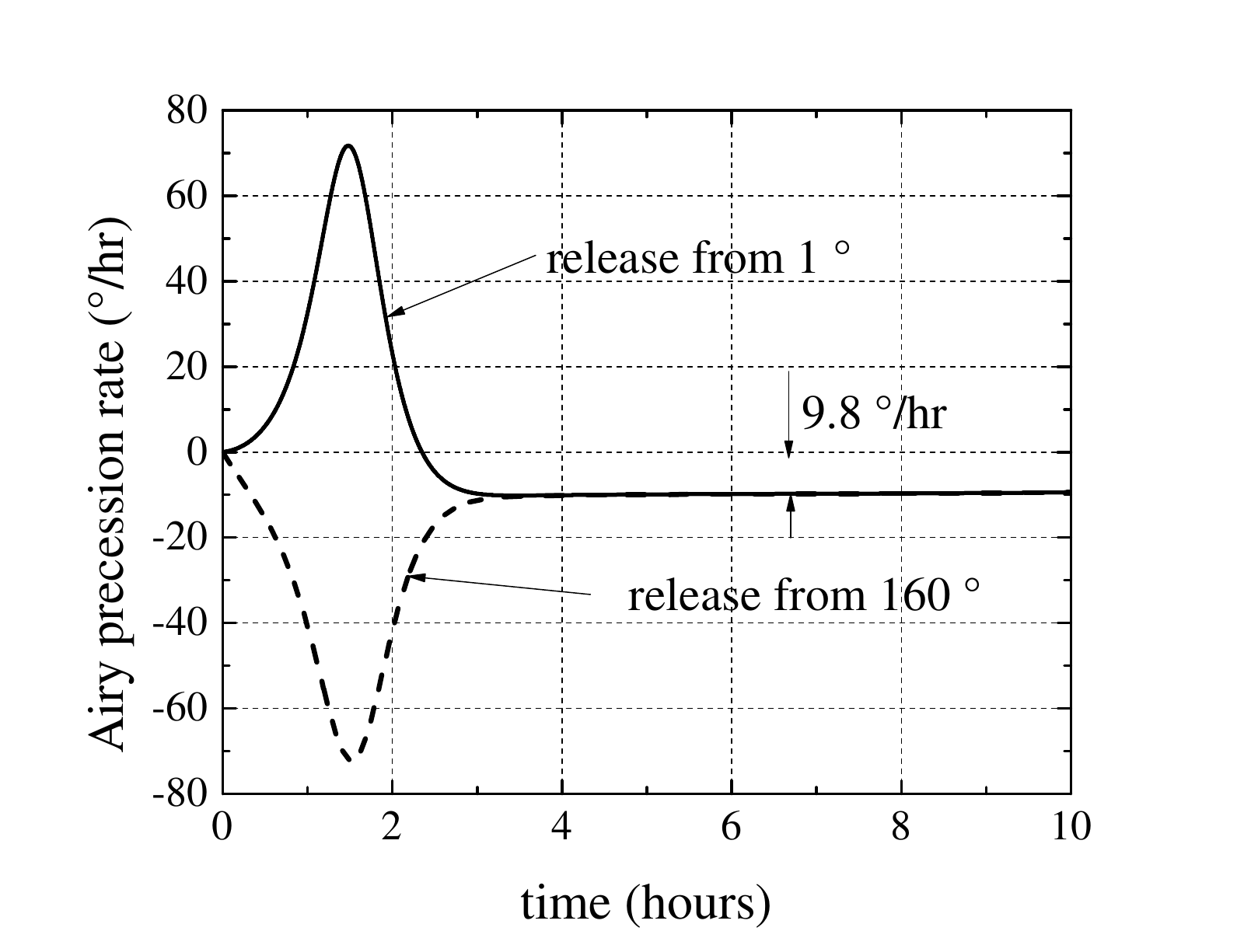} \\  
    (c) \hspace*{7cm} (d) \\
    \caption{Pendulum with $L= 480.3$ cm, $T_x=4.3946$ s, $T_y=4.3948$ s, $\Delta T= 2.2 \cdot 10^{-5}$, $\Delta L$ = 0.43 mm, and oscillation amplitude $\displaystyle \ell_0 =30$ cm. The first curve is $\varphi_1=1$ $^\circ$ and the second curve is for $\varphi_1=160$ $^\circ$.  (a) Angle of azimuth. (b) Zoom of Angle of azimuth. (c) Minor Semiaxis. (d) Airy precession rate.}
    \label{fig:3}
\end{figure}


Our final example, illustrated in Fig. \ref{fig:3}, demonstrates the behavior of a pendulum released in a quadrant where \Apr{} is negative. Initially, the pendulum precesses CCW due to Coriolis precession. However, over time, the rate of precession gradually decreases until it eventually stops. At the outset, when $t=0$, the Airy precession is zero. As time progresses, the Airy precession rate increases until it becomes significant enough to overcome Coriolis precession. This leads to an immediate reversal in direction (at 0.36 hours, as shown in Fig. \ref{fig:3}(b)), resulting in CW precession. Airy precession continues to increase until the precession stops again in an angle of the same quadrant, where $\Apr{}=-\Cpr{}$. This stopping angle corresponds to the angle observed in experiments when the bob is released from $\varphi_1=1 ^\circ$.

In Fig. \ref{fig:3}, we plot the azimuth angle $\varphi$, the minor semiaxis, and the Airy precession rate for pendulums launched from $\varphi_1=1 ^\circ$ and $\varphi_1=$160 $^\circ$. Notably, the stopping angle is identical for both launch points: $\varphi_s = 93.14^\circ$, as shown in Table I. This peculiar behavior was observed in our experiments with a pendulum of $480.3$ cm in length. Additionally, there is an equilibrium angle in the fourth quadrant, given by $\varphi_s' =\varphi_s+180^\circ$. It is important to highlight that there are no equilibrium angles perpendicular to $ \varphi_s$ or $\varphi_s'$, as these would be located in the first or third quadrants, where both Coriolis and Airy precession are positive.

In Fig. \ref{fig:3}(a), we show the behavior over 40000 iterations, which corresponds to more than 1.5 days of oscillation. We can observe that approximately every 14 hours, the angle deviates from the equilibrium angle $\varphi_s$ for a period of time and then returns to a stable azimuth. We believe that in Eq. (\ref{eq:Min}) the continuous increase in the phase difference alters the dimensions of the ellipse, affecting Airy precession and resulting in a loss of the equilibrium state. So far, we have been unable to verify this effect in experiments, even though we have monitored the azimuth for more than 60 hrs. 



Many situations can be analyzed within this scheme: for example, launching the pendulum in different quadrants or with different starting azimuth ($\varphi_1$), or to analyze very bad supports which have large values of $\Delta\omega$.

\section{Conclusions}
The most important conclusion is this: if a Foucault pendulum is built, Airy precession must be controlled in order to demonstrate or accurately measure Coriolis precession.

If you build a pendulum without any kind of restitution to keep the amplitude constant, it is necessary to find a length and an oscillation amplitude where the Airy precession rate is small enough to see Coriolis precession in a reasonable period of time. Or, you may need a pendulum support with lower anisotropy.

In our hypothetical calculations for the pendulum of 67 m, Coriolis precession could be seen for more than ten hours. If the pendulum is released from 45 $^\circ$ relative to the principal axes, this time could be longer.
If Airy precession starts to perturb the motion, the pendulum may be stopped and released again in the same azimuth, resetting the time and the phase $\Delta\omega t$, making \Apr{} restart with its minimum value for this azimuth.

There are other methods to control Airy precession.  We cite two methods from the literature. In the first, Professor R. Crane made the bob kick a ring when it arrived at its maximum amplitude, which brakes the maximum speed along the minor axis.\cite{Crane} This reduces the minor axis, and thereby reduces the Airy precession rate.

A second method, first implemented by Mastner in 1984, is to use a conductive ring.\cite{Mustner} When the bob reaches its maximum amplitude, the magnet of the bob (used to maintain constant the amplitude) interacts with the conductive ring, and because of Foucault's currents (also known as eddy currents), it brakes Airy's precession.
This method was used in Ref. \onlinecite{Salva} for a Foucault pendulum 5m long with a 10 cm oscillation amplitude with a copper ring as the conductive material. The pendulum oscillated for almost 5 years continually. The electromagnetic brake reduced the Airy precession rate  to less than $0.3 \, ^\circ/$h, while the rate of Coriolis precession was around 10 $\, ^\circ/$h. 
The force exerted by the electromagnetic brake against the movement of the minor axis is proportional to the tangential speed of the bob, the thickness and conductivity of the ring, and the magnetic flux through the conductive surface.  This brake is useful to reduce Airy precession, but will not reduce it to zero, because a velocity between the components is needed to operate. {If an electromagnetic brake is used the bob mass should be as low as possible, in contrast to the common approach of using the heaviest bob possible.\cite{Somerville,Sommeria} An analysis of how the electromagnetic brake modifies the minor and major axes of the ellipse can be found in Ref. \onlinecite{Pippard}.}

Our model could be useful to improve the design of Foucault pendulums. It can reveal the best parameters for the accurate observation of Coriolis precession: amplitude, length, weight, shape of the bob, etc. It could also be used to validate the support, an electromagnetic brake, or a design to ensure continuous oscillation.
Fig. \ref{fig:2} illustrates the usefulness of such calculations. The variations of minor semiaxis, major semiaxis, and \Apr{} repeat again and again, and precession doesn't stop. The pendulum will oscillate and precess continuously, but with enormous variation of the total precession, as in Fig. \ref{fig:2}(d). 

Another example, that doesn't account for experimental data, is to model what happens if the pendulum is hung at the equator, where there is no Coriolis force. Our model predicts that if the bob is released at a principal axis, it will continue oscillating in a linear trajectory. This is because at the principal axes there is no Airy precession, and because of the absence of Coriolis precession, the angular velocity is always zero. If the bob is released elsewhere, the angular velocity will slowly increase due to Airy precession, and it will precess until it reaches a principal axis, where it will oscillate without precessing.

One more useful feature of the model is that we can estimate the critical oscillation amplitude for which a pendulum will stop precessing or precess continuously using Eq. (\ref{eq:amp_osc}). The total precession rate will vary from zero to 2\Cpr{}, because $\Apr{} < \Cpr{}$ and the total precession rate is $\Apr{}+ \Cpr{}$. To use a pendulum to measure \Cpr{} from direct observation, we recommend an electromagnetic brake to reduce Airy precession.

\section{Acknowledgements}
We want to acknowledge Dra. Graciela Bertolino for her encouragement  for writing this paper, Dr. Diego Cuscueta and Eng. Eduardo Taglialavore for always being there when we had problems of any kind, and Tech. Carlos Talahuer for his enthusiasm and being always ready for building the beautiful elements of the pendulums. We thank Dr. Mariano Gomez Berisso for his important contribution in computer programming and experimental physics discussions. We also want to thank the staff of Instituto Balseiro, the Leo Falicov Library, and the Centro Atómico Bariloche for their contributions to this work in various ways. Finally we greatly acknowledge Prof. Rene Verreault for useful comments and for providing us important references.
\section{Data Availability Statement}
The data that support the findings of this study are available within the article.

\section{Conflict of Interest statement}
All authors declare that they have no conflicts of interest to disclose.

\bibliographystyle{unsrt}
\bibliography{Salva2025}
\pagebreak
\appendix
\section{Axes of the elliptical path}
In this appendix we derive the semiaxes of the elliptical trajectory of our motion model. We start with Eq. (\ref{eq:motion}).
Let $\delta=\Delta \omega t$. We assume that  $\delta$ is approximately constant during one period of oscillation. If this assumption does not hold, then the trajectory is not a stable ellipse. Using the trigonometric identity for addition in the cosine function, we have:

\begin{equation*} 
\begin{cases}
x = A \cos ( { \omega t})+B \sin ( { \omega t})\\
y = C { \cos (\omega t)}  
\end{cases}
\end{equation*}

where
\begin{eqnarray}
A= \ell_0 \cos(\varphi ) \cos ( \delta)\\
B= -\ell_0 \cos(\varphi ) \sin ( \delta)\\
C= \ell_0 \sin(\varphi )
\end{eqnarray}

Then we use the trigonometric identity: $\sin ( { \omega t})^2+\cos ( { \omega t})^2=1$, to obtain the following quadratic expression for $x$ and $y$:
\begin{equation} \label{eq:quadratic}  
C^2x^2+(A^2+B^2)y^2-2ACxy=B^2 C^2    
\end{equation}

Next, we apply the Principal Axes Theorem for quadratic forms to find the lengths of the major and minor axis:\cite{Kolman} \\

\noindent\makebox[\linewidth]{\rule{\textwidth}{1pt}}
\begin{teo}
Let $ex^2 + fxy + gy^2 = h$ represent a quadratic equation for the variables $x$ and $y$. Then, there exists a unique $\alpha \in [0, 2 \pi)$ such that the quadratic equation can be expressed as:
\begin{equation}
\lambda_1 v_1^2 + \lambda_2 v_2^2 = h \label{eq:teorot}    
\end{equation}

where $v_1$ and $v_2$ are the axes obtained through the rotation of the $x$ and $y$ axes by an angle $\alpha$ counterclockwise, and $\lambda_1$ and $\lambda_2$ are the eigenvalues of the following matrix:

$$ \left( \begin{array}{cc}
    e & f/2 \\
f/2 & g      
   \end{array} \right) $$
\end{teo}
\noindent\makebox[\linewidth]{\rule{\textwidth}{1pt}}
Observe that equation (\ref{eq:teorot}) is equivalent to:
$$\frac{v_1^2}{\left(\sqrt{\frac{h}{\lambda_1}}\right)^2} + \frac{v_2^2}{\left(\sqrt{\frac{h}{\lambda_2}}\right)^2}  = 1$$
Therefore, the two semiaxes of the ellipse are obtained by computing: $ \sqrt{h/\lambda_i}$.  
The matrix of the quadratic form (\ref{eq:quadratic}) is:

$$ \left( \begin{array}{ccc}
 C^2  & &-AC \\
-AC & \quad& A^2 +B^2     
   \end{array} \right) $$
which has the following eigenvalues:

$$\lambda_1= \frac{\ell_0^2}{2} \left(1 -\sqrt{1-\sin^2(2 \varphi) \sin^2(\delta)} \right), \qquad \lambda_2= \frac{\ell_0^2}{2} \left(1 + \sqrt{1-\sin^2(2 \varphi) \sin^2(\delta)} \right).$$
Direct substitution for $B, C$, and $\lambda_i$ in the semiaxes of the ellipse given by:
$$\sqrt{\frac{B^2 C^2}{\lambda_i} },$$ 
yields Eqs. (\ref{eq:Maxe}) and (\ref{eq:Min}).

\section{ Rotation of the elliptical path}

In this appendix we give a detailed description of the bob's motion in our model. 
Eq. (\ref{eq:motion0}) shows that the bob moves either counterclockwise (CCW) or clockwise (CW) around its elliptical trajectory, depending on the phase difference between the principal axes. Along the principal axes ($\varphi=0 ^\circ, 90^\circ,180 ^\circ,270 ^\circ$), the trajectory is a straight line, and every time the bob passes through one of the principal axis, the sense of rotation around its elliptical trajectory changes.

In the first quadrant ($0<\varphi<90^\circ $), the bob will rotate CCW, and in the second quadrant ($90^\circ<\varphi<180^\circ $), the bob will rotate CW. In the third quadrant, the bob will rotate CCW and in the fourth quadrant CW.

\begin{figure}[h!]
 \includegraphics[height=6cm,clip=true,trim=0 0 0 0]{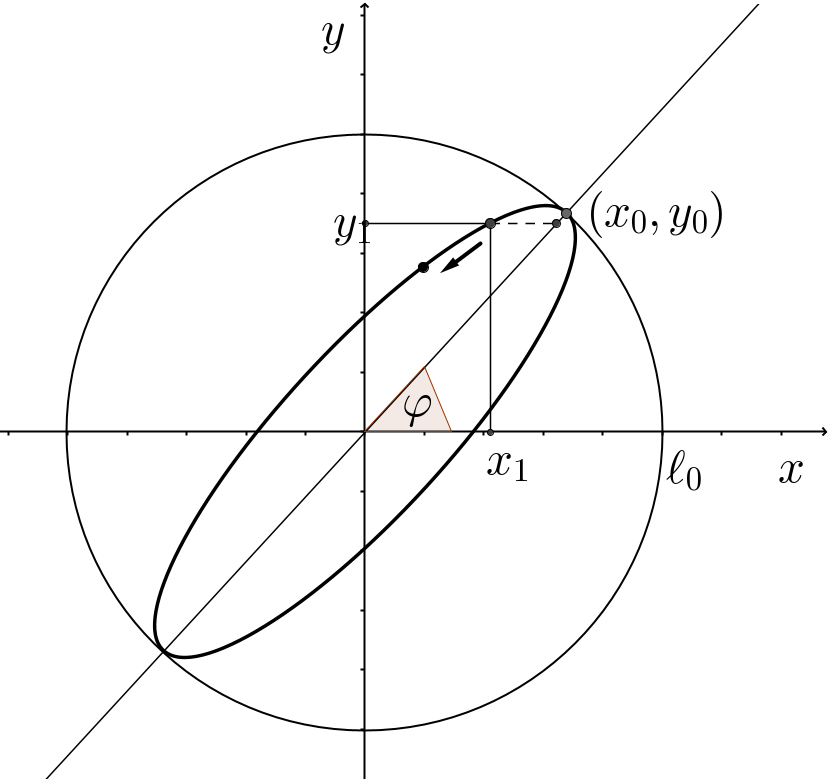}
 \caption{Sense of rotation of the ellipse} \label{fig:elipseCW}
\end{figure}

In Fig.~\ref{fig:elipseCW} we show the ellipse of the model, for $\Delta \omega \approx 0$ and positive. Note that for $t_0=0$, we have $(x_0,y_0)=(\ell_0\cdot\cos(\varphi),\ell_0\cdot\sin(\varphi))$: $(x_0,y_0)$ is a point in the circle of radius $\ell_0$, situated at $\varphi$. For a later time $t_1$ close to $t_0$, the cosine function is decreasing, and $\cos(\omega t_1)<\cos(\omega t_1+\Delta \omega t_1)$. Therefore $x_1=\ell_0\cdot\cos(\varphi)\cos(\omega t_1+\Delta \omega t_1) < \ell_0\cdot\cos(\varphi)\cos(\omega t_1)$, that is the point $(x_1,y_1)$ is located to the left of $(\ell_0.\cos(\varphi)\cos(\omega t_1),\ell_0.\sin(\varphi)\cos(\omega t_1))$. This implies that the point $(x_1,y_1)$ is above the line $y=x \, \tan(\varphi)$. Therefore, the ellipse will move CCW when $\varphi$ is in the first quadrant.

On the other hand, if $\Delta \omega \approx 0$ but negative, then $\cos(\omega t_1+\Delta \omega t_1)>\cos(\omega t_1)$, and the point $(x_1,y_1)$ will be below the line $y=x \, \tan(\varphi)$. Therefore, the bob will move CW when $\varphi$ is in the first quadrant.

Thus, the phase $\Delta \omega t$ between the axes determines the sense of rotation.

\section{Brief description of the experimental pendulum} \label{secc:pend}

\begin{figure}[ht!]
    \centering
  \includegraphics[height=5.5cm,clip=true, trim=0 0 0 0]{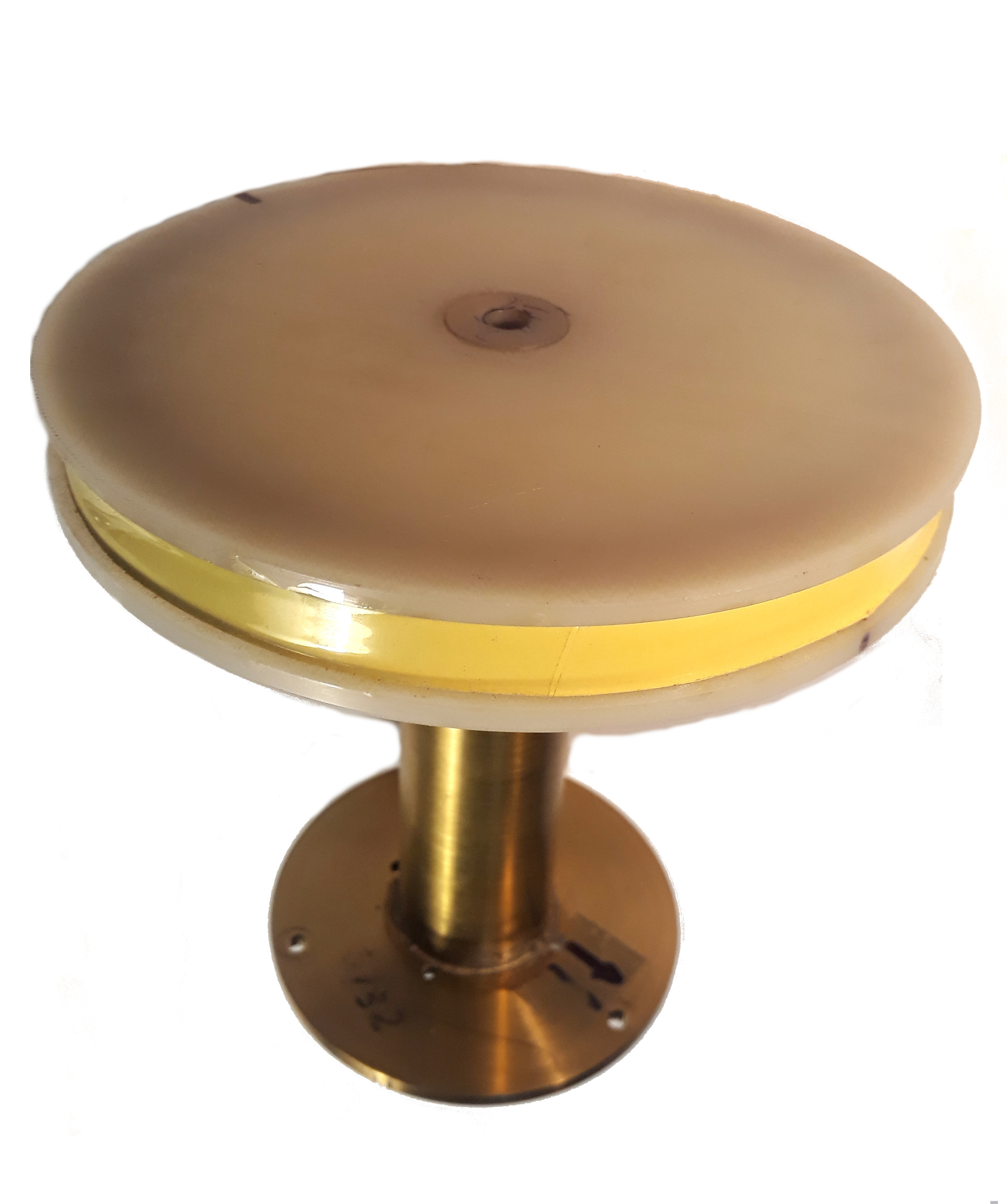}   \includegraphics[height=6.5cm, clip=true,  trim=0 0 0 0,angle=0,origin=c]{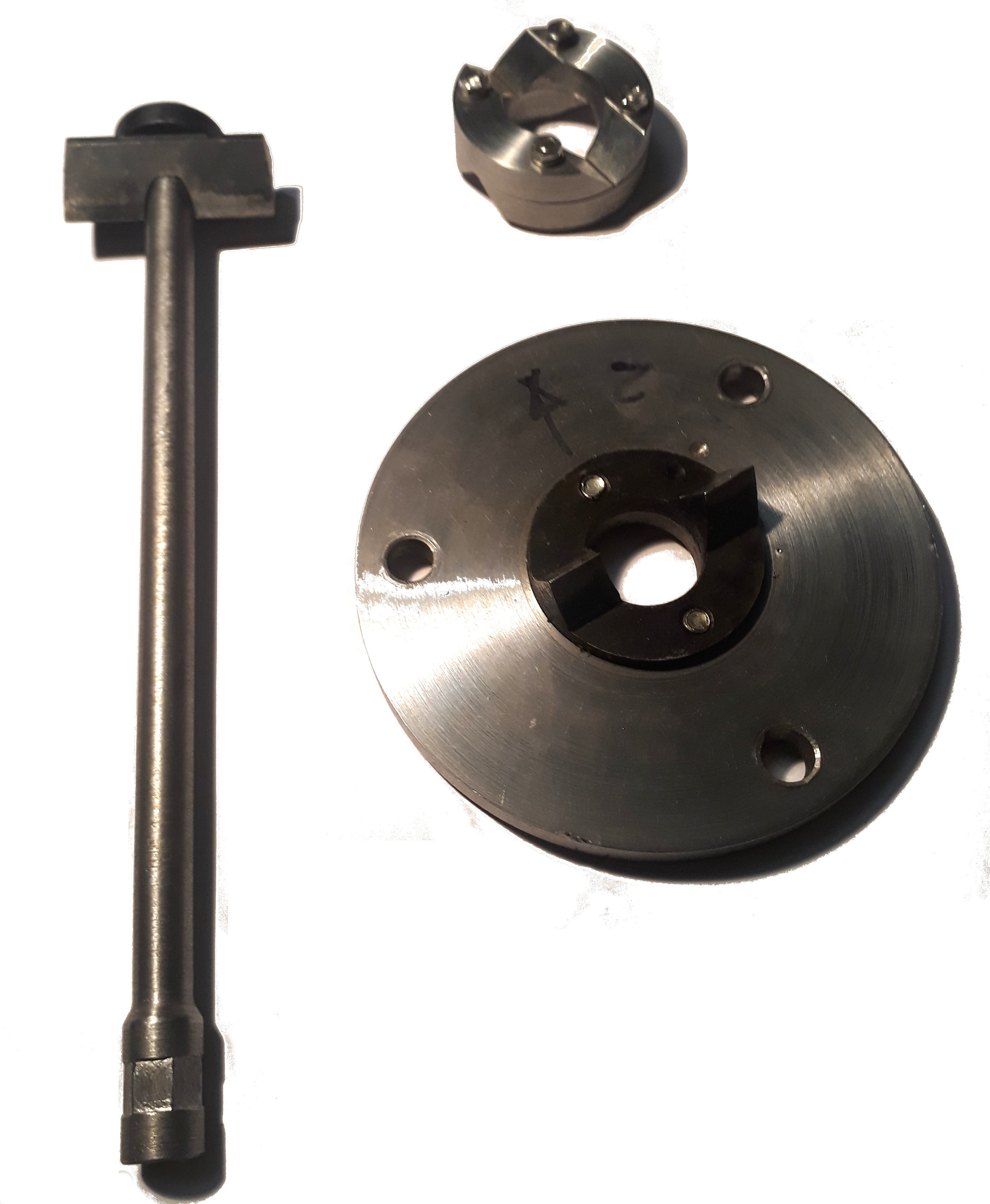} \\
    (a) \hspace*{7cm} (b) \\
    
    \caption{(a) Impulsion coil and synchronization coil (small). (b) Support.}
    \label{fig:fotopend}
\end{figure}

\begin{figure}[ht!]
    \centering
         \includegraphics[height=4cm,clip=true, trim=0 0 0 0]{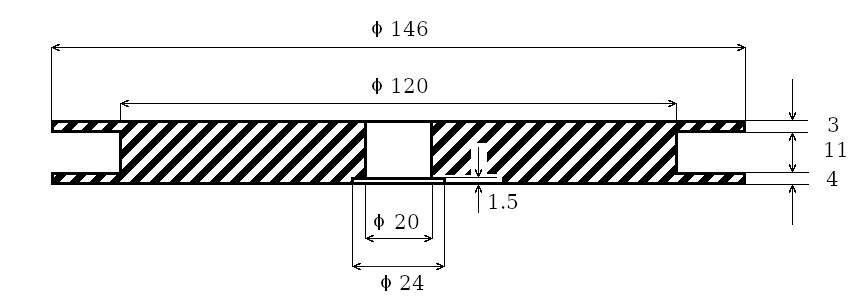}   \includegraphics[height=4cm, clip=true,  trim=0 0 0 0,angle=0,origin=c]{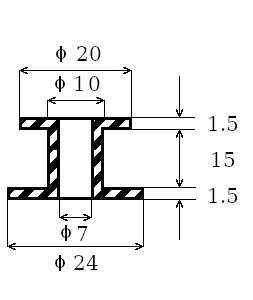}\\
     \hspace*{3cm}(a) \hspace*{6.5cm} (b) \\
    \includegraphics[height=8cm,clip=true, trim=0 0 0 0]{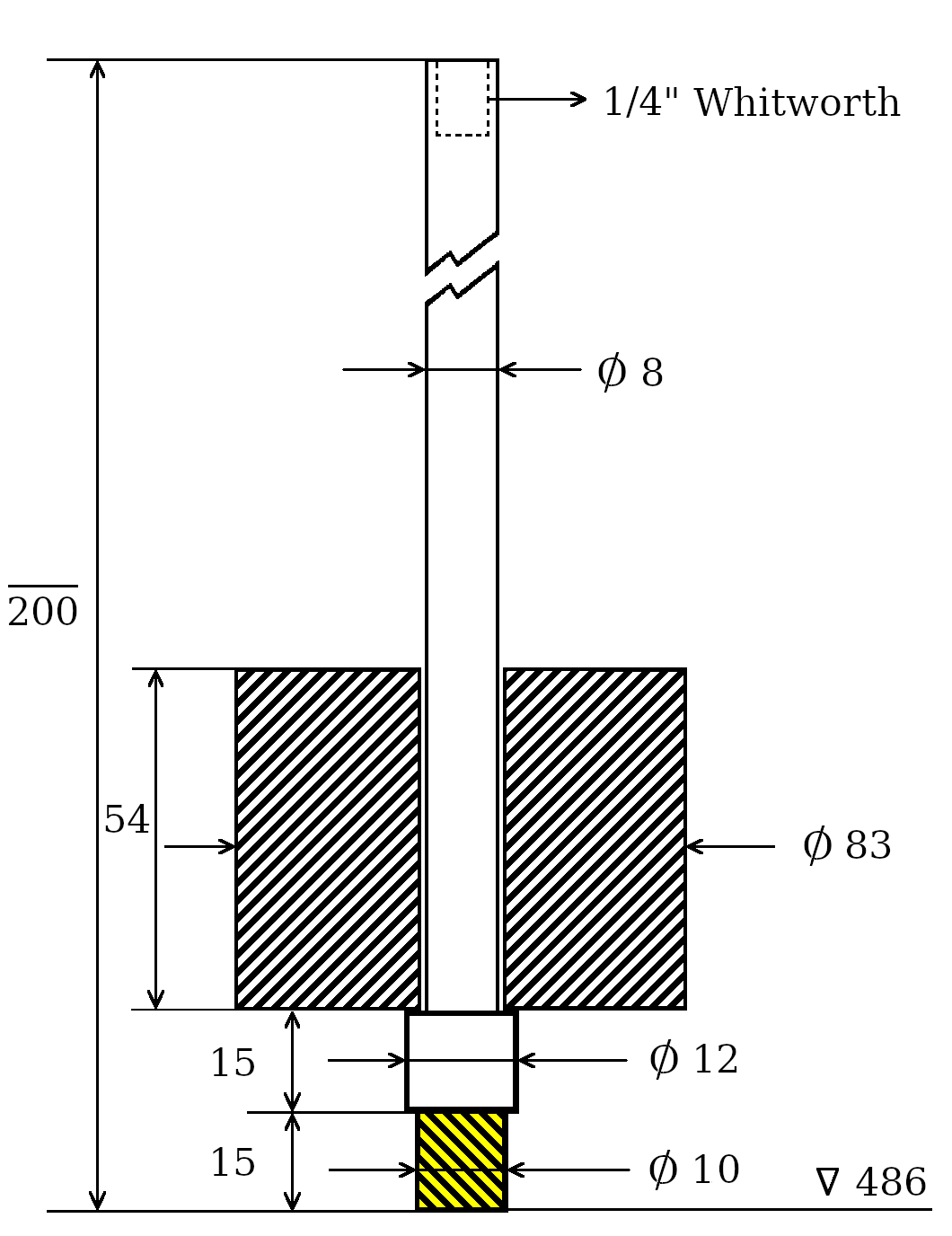} \\
    (c)
    \caption{Schematic diagrams of: (a) the impulsion coil, (b) the synchronization coil, (c) the bob. The units are millimeters.}
    \label{fig:planos}
\end{figure}

The support of the experimental pendulum was a gimbal with a metallic crosshead (see Fig. \ref{fig:fotopend}(b)). The stabilized amplitude system comprised an impulsion coil under the base and a synchronization coil (see Fig. \ref{fig:fotopend}(a)), an electronic circuit that operates the coils, and a magnet located at the bottom of the bob (see Fig. \ref{fig:planos}(c)).  The impulse coil is meticulously wound, layer by layer, with insulated copper wire of 0.65 mm in diameter, so that its field can be approximated by that of an ideal coil. The synchronization coil is located inside the impulsion coil, and is wound with insulated copper wire of 0.1 mm in diameter, filling the available volume completely.

In Fig. \ref{fig:planos}, we present the diagrams of the two coils and of the bob. The bob is a cylinder made of lead that weights 2870 g, while the support is crafted from bronze. A rare-earth magnet is positioned at the bottom, penetrating 10 mm into the bob. It is cylindrical in shape, measuring 25 mm in length and 10 mm in diameter. The upper part of the support is attached to a screw that is connected to the wire. This design of the bob's support allows us to easily change the bob itself and experiment with heavier ones. 

We recommend using non-magnetic materials, such as lead or bronze, to avoid potential issues related to the stabilized amplitude system.

\end{document}